\documentclass[aip, jcp, notitlepage, floatfix, reprint, citeautoscript, twocolumn]{revtex4-1}

\usepackage{amsmath}
\usepackage{amsfonts}
\usepackage{graphicx}
\usepackage{dcolumn}
\usepackage[table]{xcolor}
\usepackage[version=3]{mhchem}
\usepackage{transparent}
\usepackage{gensymb}
\usepackage{tabularx}
\usepackage{etoolbox}
\usepackage[hidelinks]{hyperref}
\usepackage{bm}
\usepackage{relsize}
\usepackage[normalem]{ulem}
\usepackage{bbm}

\def\maxfloatwidth{%
  \ifdim\columnwidth>246.0pt
  300.0pt  \else
  \columnwidth
  \fi
}

\newcommand{\tbf}[1]{\textbf{#1}}

\newcommand{\mrm}[1]{\mathrm{#1}}
\newcommand{\mbf}[1]{\mathbf{#1}}
\newcommand{\tcr}[1]{\textcolor{black}{#1}}

\newcommand{\etal}{\emph{et al.}}

\newcommand{\dctfac}{\bigg(\frac{\epsilon-1}{\epsilon}\bigg)}

\begin{document}

\title{Quadrupole-mediated dielectric response and the charge-asymmetric solvation of ions in water}

\author{Stephen J. Cox}
\affiliation{Yusuf Hamied Department of Chemistry, University of
  Cambridge, Lensfield Road, Cambridge CB2 1EW, United Kingdom}
\email{sjc236@cam.ac.uk}

\author{Kranthi K. Mandadapu} 
\affiliation{Chemical Sciences Division, Lawrence Berkeley National
  Laboratory, Berkeley, CA 94720, United States.}
\affiliation{Department of Chemical and Biomolecular Engineering,
  University of California, Berkeley, CA 94720, United States.}
\email{kranthi@berkeley.edu}

\author{Phillip L. Geissler}
\affiliation{Chemical Sciences Division, Lawrence Berkeley National
  Laboratory, Berkeley, CA 94720, United States.}
\affiliation{Department of Chemistry, University of California,
  Berkeley, CA 94720, United States.}
\email{geissler@berkeley.edu}

\date{\today}

\begin{abstract}
  Treating water as a linearly responding dielectric continuum on
  molecular length scales allows very simple estimates of solvation
  structure and thermodynamics for charged and polar solutes. While
  this approach can successfully account for basic length and energy
  scales of ion solvation, computer simulations indicate not only its
  quantitative inaccuracies but also its inability to capture some
  basic and important aspects of microscopic polarization
  response. Here we consider one such shortcoming, a failure to
  distinguish the solvation thermodynamics of cations from that of
  otherwise-identical anions, and we pursue a simple, physically
  inspired modification of the dielectric continuum model to address
  it. The adaptation is motivated by analyzing the orientational
  response of an isolated water molecule whose dipole is rigidly
  constrained. Its free energy suggests a Hamiltonian for
  dipole
  fluctuations that accounts implicitly for the influence of
  higher-order multipole moments, while respecting constraints of
  molecular geometry. We propose a field theory with the suggested
  form, whose nonlinear response breaks the charge symmetry of ion
  solvation. An approximate variational solution of this theory, with
  a single adjustable parameter, yields solvation free energies that
  agree closely with simulation results over a considerable range of
  solute size and charge.
\end{abstract}

\maketitle

\section{Introduction}
\label{sec:intro}

Water is perhaps the most important solvent, and understanding the
fundamental physical principles that underlie aqueous solvation is
essential to a broad range of disciplines, such as protein structure
and dynamics, desalination, atmospheric chemistry and
crystallization. Despite numerous studies over the past century, major
gaps in our understanding of aqueous solvation still exist,
particularly for small, charged solutes and for environments that are
spatially heterogeneous.  Highlighting these gaps, active research
continues to develop and apply increasingly sophisticated methods of
spectroscopy
\cite{tielrooij2010cooperativity,OttenSaykally2012sjc,verreault2012conventional,piatkowski2014extreme,mccaffrey2017mechanism,chen2016electrolytes}
and computer simulation
\cite{HummerGarcia1998sjc,LumWeeks1999sjc,ashbaugh2000convergence,horinek2007specific,levin2009polarizable,ben2016interfacial,beck2013influence,remsing2014role,cox2020assessing}
in order to clarify the solvation of ions in aqueous systems.

These limits on our understanding are reflected by the lack of a
robust, general, and thoroughly predictive theory for the microscopic
structure and thermodynamics of water's response to charged
solutes. As a promising and historically significant starting point,
dielectric continuum theory (DCT) -- a macroscopic linear response
theory for solvent polarization -- can be applied in a microscopic
context. This approach has yielded insights that inspire modern
perspectives on solvation, but its flaws and limitations are
considerable.  Among the most straightforward and important
microscopic applications of DCT, the Born model of solvation
\cite{born1920volumen} caricatures an ionic solute as a
volume-excluding, uniformly charged sphere of radius $R$ embedded in a
continuous, linearly-responding solvent medium with dielectric
constant $\epsilon$.  Reversibly introducing the solute's charge $q$
in this model gives a change in free energy
\begin{equation}
  \label{eqn:born-plain}
  F_{\rm chg}^{(\rm Born)}(q) = -\frac{q^2}{2R}\dctfac
\end{equation}
that explains the basic energy scale of ion solvation and its
sensitivity to solute size and charge, and asserts the permittivity as
an essential determinant of solvent quality.  Its quantitative
predictions are roughly correct, provided that the dielectric radius
$R$ is treated as an empirical parameter similar but not identical to
the radius $R_0$ of molecular volume exclusion.

\begin{figure}
  \includegraphics[width=7.65cm]{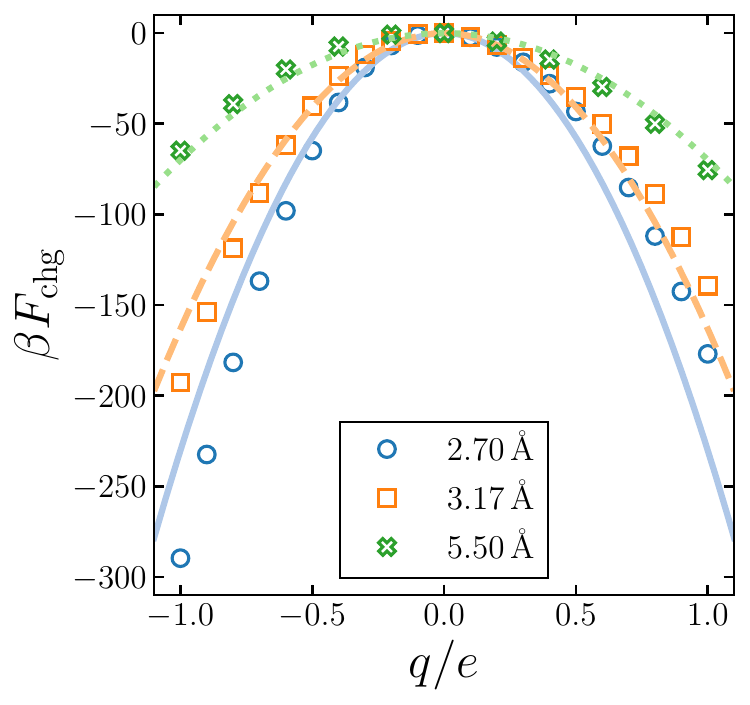}
  \caption{Solute charging free energy $F_{\rm chg}$ vs $q$ for
    different solute sizes $R_0$, as indicated in the legend.
    Symbols
    show results from simulations, while lines show best-fits of
    $F_{\rm chg}^{\rm (Born)}(q)$ (Eq.~\ref{eqn:born-plain}) to
    $F_{\rm chg}$. $F_{\rm chg}^{\rm (Born)}(q)$ largely captures the
    overall scale and size dependence of $F_{\rm chg}$, but it does
    not describe the asymmetric solvation
    \cite{latimer1939free,rajamani2004size,bardhan2012affine} of
    anions vs cations.
  }
  \label{fig:fe_born}
\end{figure}

Fig.~\ref{fig:fe_born} shows the Born estimate of the charging free
energy $F_{\rm chg}(q)$, alongside results of molecular simulation of
the SPC/E model of water \cite{BerendsenStraatsma1987sjc}, as a
function of solute charge for several solute sizes $R_0$. (As a
  measure of $R_0$ in molecular simulations, we quote values of the
  Lennard-Jones diameter for solute-water interactions, which is a
  reasonable---though not unique---choice for neutral solutes
  \cite{HANSEN2013265}. Effective hard sphere radii for fully charged
  ions, as judged from radial distribution functions, are
  approximately 10-20\% smaller. This convention is used throughout
  the paper.)  Relative to $F_{\rm chg}(e)$, where $e$ is the charge
of an electron, agreement is reasonable even for ions as small as
fluoride. But absolute errors of $\sim 50 k_{\rm B}T$ (where $T$ is
the temperature and $k_{\rm B}$ is Boltzmann's constant) overwhelm the
scale of typical thermal fluctuations.  Despite the magnitude of these
errors, DCT continues to serve as a basis for quantifying the
thermodynamics of aqueous response in calculations that cannot afford
to represent water molecules explicitly \cite{tomasi2005quantum}.
Motivated in part by such usage, this paper describes theoretical
efforts to improve on DCT while maintaining the simplicity underlying
its appeal.

From a molecular perspective, DCT is remarkably undetailed, resolving
neither the tetrahedral motifs defining water's hydrogen bond network
nor the features of molecular geometry that are responsible for
it. Significant improvement might well require an approach that
differs substantially in both spirit and methodology. Indeed, there is
ample evidence that near-field contributions from a solute's immediate
environment have a different character than contributions from more
distant molecules. The idea that the latter are well described by DCT,
while the former are not, has figured prominently in many theoretical
and computational approaches to describing aqueous environments.  The
inner-shell of Marcus theory \cite{marcus1956theoryI}, for example,
acknowledges and empirically addresses such a distinction in the
nature of near-field and far-field response. More directly relevant to
our study, the multi-state Gaussian model \cite{HummerGarcia1997sjc}
of single ion solvation similarly presumes that local
solvent-structure requires special treatment, while far-field response
obeys simple Gaussian statistics.  A precise and systematic
computational framework for range separation in solvation can be found
in the quasi-chemical theory (QCT)
\cite{pratt1998quasi,HummerGarcia1998sjc,2012TPDT} developed by Pratt
and co-workers. These hybrid theories and methodologies can achieve a
high degree of accuracy, e.g. when QCT is used in combination with
\emph{ab initio} treatment of electronic structure
\cite{duignan2017real,duignan2017electrostatic}.  But they are not
nearly as flexible or generalizable as DCT.

In the next section we describe in detail the specific and fundamental
shortcoming of DCT that inspires our theoretical development, namely,
an inability to distinguish between the solvation of cations and
anions that differ only in the sign of their charge. The contrastingly
strong charge asymmetry observed in molecular simulations is then
framed in terms of a water molecule's higher-order multipole moments,
with particular emphasis on the molecular
quadrupole. Sec.~\ref{sec:DevelopingFieldTheory} shows, in the context
of a single water molecule, how integrating out quadrupole
fluctuations renormalizes statistics of the molecular dipole. The
result of that molecular calculation is then used to motivate a
generalized version of DCT, whose predictions for ion solvation
thermodynamics are approximately explored and numerically evaluated in
Sec.~\ref{sec:Assessing}. We end with a discussion and outlook in
Sec.~\ref{sec:Outlook}.

\section{Background theory and simulation}
\label{sec:BackgroundTheory}

\subsection{Charge asymmetric solvation}
\label{subsec:AsymSolv}

This paper focuses on one key failing of DCT, evident in
Fig.~\ref{fig:fe_born} for the solvation of ions in bulk liquid water.
Specifically, in molecular simulations cations and anions of the same
size can have very different solubilities, while DCT in its simplest
form lacks such charge asymmetry completely. Interfacial solvation
provides even more striking examples of charge asymmetry, with some
anions adsorbing favorably to the liquid's outermost layer while their
cationic counterparts are strongly depleted. Here we will consider
only the bulk case.

The lack of charge aymmetry in DCT can be readily appreciated from its
basic mathematical structure. As a linear response theory, DCT in its
simplest form can be cast as a microscopic model for a Gaussian
fluctuating
dipole
field ${\bf m}_{\bf r}$, with energy
\cite{song1996gaussian}
\begin{widetext}
\begin{equation}
  \label{eqn:dct-mic}
       {\cal H}_{\rm dip}[{\bf m}_{\bf r}] =
       \frac{1}{2}
\sum_{\bf r}\sum_{{\bf r}'} {\bf m}_{\bf r}\cdot
       \left[
         \bigg(\frac{4\pi}{\epsilon-1}\bigg)
         v^{-1}\,
              {\bf I}\, \delta({\bf r} - {\bf r}')
              +
              \nabla\nabla' \frac{1}{|{\bf r}-{\bf r}'|}\right]
       \cdot {\bf m}_{{\bf r}'} 
\end{equation}
\end{widetext}
where ${\bf I}$ is the identity tensor.  This model can be
equivalently formulated in continuous space
\cite{MaddenKivelson1984sjc}, but it will later become more convenient
for us to view space discretely. We therefore take the position vector
${\bf r}$ to index a lattice cell with microscopic volume $v$ and
${\bf m}_{\bf r}$ to be a coarse-grained representation of the
molecular dipole distribution within $v$. The coarse-graining
transformation could take many forms, and we will not specify one
here. Introducing a solute with charge $q$, which we place at the
origin without loss of generality, adds an electrostatic interaction
between ${\bf m_r}$ and the solute's electric field ${\bf E}_q({\bf
  r}) = -q \nabla r^{-1}$,
\begin{equation}
  \label{eqn:dct-mic-int}
        {\cal H}_{\rm DCT}[{\bf m}_{\bf r}] =
        {\cal H}_{\rm dip}[{\bf m}_{\bf r}]
        - \sum_{\bf r}\, {\bf E}_q({\bf r}) \cdot {\bf m}_{\bf r}
\end{equation}
Volume exclusion is acknowledged in this description only by
restricting the sums in Eqs.~\ref{eqn:dct-mic}
and~\ref{eqn:dct-mic-int} to lattice cells that are not occupied by
the solute, a restriction we will leave implicit.  The thermodynamic
consequences of evacuating the solute's volume, while significant in
some cases, are not considered by DCT and will not be accounted for
here.  Our focus on comparing cations and anions of the same size
justifies this neglect, but is not meant to minimize the complex and
interesting coupling between density and polarization fields, which is
particularly important near interfaces
\cite{ballenegger2005dielectric,schlaich2016water,loche2018breakdown,zhang2020electromechanics}.

\begin{figure}
  \includegraphics[width=7.65cm]{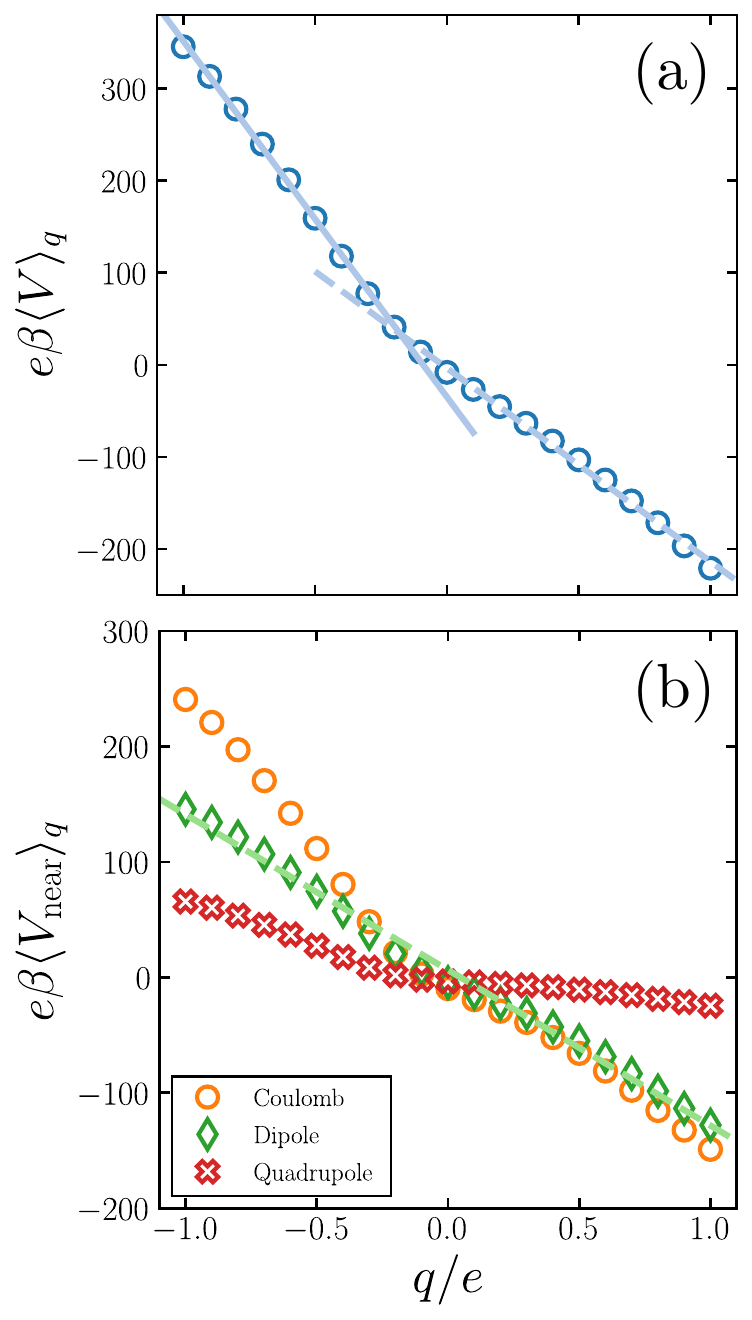}
  \caption{Average electrostatic potential at the center of a
    volume-excluding solute, with size $R_0=3.17$\,\AA{} and charge
    $q$, due to the surrounding solvent.  (a) All solvent molecules in
    the simulation are included (see Sec.~\ref{sec:Methods}). The
    solid and dashed lines are guides to the eye, suggesting distinct
    susceptibilities for $q/e < -0.2$ and $q/e > -0.2$,
    respectively. (b) Average electrostatic potential at the center of
    the solute due to molecules in the first coordination shell (those
    molecules with oxygen atoms within 3.5\,\AA{} of the solute's
    center). Along with the full Coulomb potential, the dipolar and
    quadrupolar contributions are also shown.  The dashed green line
    is a guide to the eye, suggesting that the dipolar response is
    approximately linear.  The quadrupolar response, by contrast,
    exhibits a `kink' similar to that of the full Coulomb potential.}
  \label{fig:phi_plots}
\end{figure}

The field theoretic Hamiltonian in Eq.~\ref{eqn:dct-mic-int} is {\em
  charge symmetric}: Changing the sign of $q$, while simultaneously
inverting the dipole field's orientation, leaves ${\cal H}$
invariant. Cation and anion solvation are thus statistically
equivalent at this level of theory. The polarization field induced by
a cation, under inversion, is identical to that induced by an anion of
the same size.  Predicted solubilities of the two ions are equal as a
result, as is clear from the Born energy $F_{\rm chg}^{(\rm Born)}(q)$
(Eq.~\ref{eqn:born-plain}) as an even function of $q$. By contrast,
computer simulations indicate that ion solvation in water is
significantly charge {\em asymmetric}. In their seminal study of
single ion solvation using molecular simulation, Hummer
\etal\cite{HummerGarcia1996sjc} clarified this charge asymmetry by
presenting the average electrostatic potential $\langle V\rangle_q$ at
the center of a volume-excluding solute as a function of $q$. For
reference, we recapitulate those results in Fig.~\ref{fig:phi_plots}a
for a solute with $R_0=3.17$\,\AA{} immersed in SPC/E water.  Notably,
when referenced appropriately to vapor (see Eq.~\ref{eqn:Fmacro}),
this potential is negative even in the case of a neutral solute,
$q=0$, giving an impression that liquid water is intrinsically more
hospitable to cations than to anions. The physical origins of this
neutral cavity potential
\cite{ashbaugh2000convergence,rajamani2004size,bardhan2012affine}
$\langle V\rangle_0$ are surprisingly challenging to identify
precisely; profound ambiguities plague any attempt just to separate
contributions of solvent molecules near the solute and those of a
distant interface
\cite{aaqvist1998analysis,harder2008origin,arslanargin2012free,beck2013influence,horvath2013vapor,remsing2014role,remsing2016role,doyle2019importance,cox2020assessing}. Putting
aside the lack of a clear physical interpretation, the effects of a
nonzero neutral cavity potential are straightforward to include in the
DCT framework. Adding an interaction between the solute and this
innate potential,
\begin{equation}
  \label{dct-mic-int2}
        {\cal H}_{\rm DCT}[{\bf m_r}; \langle V\rangle_0]
        = {\cal H}_{\rm dip}[{\bf m_r}] - \sum_{\bf r}\, {\bf E}_q({\bf r}) \cdot {\bf m}_{\bf r} + q\langle V\rangle_0,
\end{equation}
gives a simply modified solvation energy
\begin{equation}
  \label{eqn:born-qV0}
  F_{\rm chg}^{(\rm Born)}(q; \langle V\rangle_0) = -\frac{q^2}{2R}\dctfac + q \langle V\rangle_0.  
\end{equation}
Modifying DCT in this simple way has little impact, however, on the
predicted charging free energy, at least on the scale shown in
Fig.~\ref{fig:fe_born}.  Furthermore, simulation results for $F_{\rm
  chg}(e)-F_{\rm chg}(-e)$, which compares the solubilities of fully
charged ions, indicate charge asymmetry in the direction opposite to
$\langle V\rangle_0$, favoring solvation of anions over cations.  Any
significant improvements obtained by including the neutral cavity
potential are limited to small values of $q$
\cite{rajamani2004size,bardhan2012affine}, as shown in the Supporting
Information (SI).

Deviations from charge symmetry in SPC/E water (and similar models)
are not at all limited to an offset in the solvent's electric
potential. The polarization {\em response} to charging a solute is
also distinct for cations and anions, with more substantial
consequences
\cite{HummerGarcia1996sjc,lynden1997hydrophobic,bardhan2012affine,rajamani2004size,cox2020assessing}. Fig.~\ref{fig:phi_plots}a
highlights this asymmetric response, which manifests in $\langle
V\rangle_q$ as a nonlinear dependence on $q$. Amending DCT to account
for this nonlinear response is much more challenging than introducing
a background potential $\langle V\rangle_0$.  Distinct thermodynamics
for charging cations and anions can be engineered by using different
dielectric radii in the Born model, as Latimer, Pitzer, and Slansky
pursued with quantitative success\cite{latimer1939free}.  More nuanced
empirical approaches have been based on the approximately
piecewise-linear character of $\langle V\rangle_q$, in essence
asserting different values of $\epsilon$ in different ranges of $q$
\cite{bardhan2012affine}. These {\em ad hoc} descriptions of ion
solvation, however, fall short of the flexible field theory we are
seeking. Such a theory would instead feature a microscopic Hamiltonian
that is anharmonic in the dipole field ${\bf m}_{\bf r}$, generating
distinct response to solute fields ${\bf E}_q$ with opposite signs of
$q$ as an emergent behavior.  Below we will propose a theory of this
form, motivated directly by the molecular fluctuations underlying
polarization response.

\subsection{Multipole expansion of the solvation potential}
\label{subsec:MolcMultipoles}

Charge asymmetric response in liquid water is rooted in the
inequivalent distribution of positive and negative charge within each
individual water molecule. A large molecular dipole is a
characteristic feature of this distribution, but by itself it is a
highly incomplete description. Electrostatic forces that underlie
hydrogen bonding and charge asymmetry are instead encoded in higher
order multipoles. Although a detailed description of these forces
requires a multipole expansion to high order, we will argue that a
low-order expansion may in fact be sufficient to correct qualitative
flaws of DCT.

In order to isolate important sources of nonlinear polarization in
computer simulations, we decompose the solvation potential $\langle
V\rangle_q$ according to multipole moment and distance from the
solute.  Anticipating that deviations from linear response are
dominated by the near-field environment, we show in
Fig.~\ref{fig:phi_plots}b contributions to $\langle V\rangle_q$ from
molecular dipoles and quadrupoles of water molecules in the solute's
first solvation shell. The nonlinear shape of the total near-field
contribution $\langle V_{\rm near}\rangle_q$ indeed strongly resembles
the full potential $\langle V\rangle_q$.  By contrast, contributions
from more distant molecules, presented in the SI, depend linearly on
$q$ to a good approximation, confirming expectations from previous
work that DCT accurately portrays polarization response on length
scales beyond $\sim 1$\,nm
\cite{HummerGarcia1996sjc,FigueiridoLevy1995sjc,HunenbergerMcCammon1999sjc,cox2018interfacial,cox2020dielectric}.

The dipole contribution to $\langle V_{\rm near}\rangle_q$ for a
solute at position ${\bf r}$ is defined as
\begin{equation}
  \langle V_{\rm near}^{\rm dip}\rangle_q =
  -\bigg\langle
  \sum_j h_{{\rm near},j}{\bm\mu}_j \cdot
  \nabla\frac{1}{|{\bf R}_j - {\bf r}|}
  \bigg\rangle
\end{equation}
where
\begin{equation}
  {\bm\mu}_j = \sum_\alpha q_\alpha {\bf r}_{j\alpha}
\end{equation}
is the net dipole of the $j^{\rm th}$ molecule, whose center resides
at ${\bf R}_j$. ${\bf r}_{j\alpha}$ is the position of site $\alpha$
on molecule $j$, whose charge is $q_\alpha$. The characteristic
function $h_{{\rm near},j}$ is unity if the oxygen atom of solvent
molecule $j$ resides in the solute's first solvation shell; otherwise,
it vanishes. Compared with $\langle V_{\rm near}\rangle_q$ and
$\langle V\rangle_q$, the first-shell dipolar potential $\langle
V_{\rm near}^{\rm dip}\rangle_q$ is a remarkably linear function of
$q$.  At large values of $q$, we expect significant nonlinearity in
$\langle V_{\rm near}^{\rm dip}\rangle_q$ due to dielectric
saturation, but for $|q|\leq e$ such effects are barely apparent on
the scale of Fig.~\ref{fig:phi_plots}.

By contrast, the quadrupolar contribution to $\langle V_{\rm
  near}\rangle_q$ exhibits a nonlinearity quite similar to that of the
full potential.  We define this contribution as
\begin{equation}
  \langle V_{\rm near}^{\rm quad}\rangle_q =
  \frac{1}{2}\bigg\langle \sum_j h_{{\rm near},j}
          {\mathbf{K}}_j:\nabla\nabla\frac{1}{|{\bf R}_j - {\bf r}|}
  \bigg\rangle,
\end{equation}
where
\begin{equation}
  \label{eqn:Qmol}
        {\mathbf{K}}_j = \sum_\alpha q_\alpha ({\bf r}_{j\alpha}-{\bf R}_j)
        ({\bf r}_{j\alpha}-{\bf R}_j)
  + ({\rm const}) {\bf I}
\end{equation}
is the net quadrupole of the $j^{\rm th}$ molecule. The coefficient
multiplying the identity tensor ${\bf I}$ in Eq.~\ref{eqn:Qmol} is
completely arbitrary, since ${\bf I}:\nabla \nabla |{\bf R}_j - {\bf
  r}|^{-1} = -4\pi \delta({\bf R}_j - {\bf r})$ and ${\bf R}_j$
necessarily lies outside the solute.  We will exploit this
arbitrariness below, freely adding and removing isotropic
contributions to ${\mathbf{K}}$ for convenience. Similar liberties can
be taken with higher order multipoles.

\begin{figure*}[tb]
  \includegraphics[width=16.08cm]{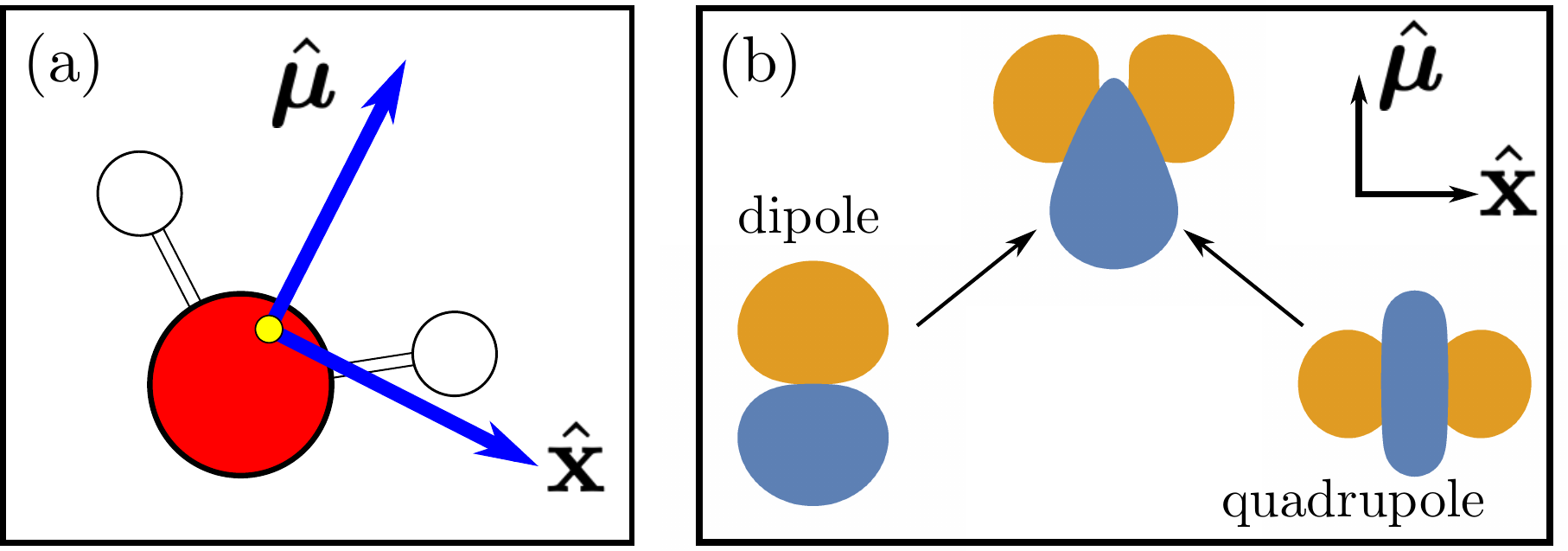}
  \caption{The molecular structure of the water molecule gives rise to
    charge asymmetry. (a) The unit separation vector $\hat{\mbf{x}}$
    between the two hydrogen atoms is orthogonal to the unit dipole
    vector $\hat{\bm\mu}$. The yellow circle represents the multipole
    expansion point, i.e., the molecule's center of charge. (b)
    Equipotential surfaces of the molecular dipole and quadrupole.
    Summing contributions of these two lowest order moments (as
    indicated by the arrows and detailed in Eq.~\ref{eqn:phi2}) is
    sufficient to generate a charge-asymmetric equipotential surface
    suggestive of a water molecule.}
  \label{fig:water_frame}
\end{figure*}

Parsing $\langle V_{\rm near}\rangle_q$ as in
Fig.~\ref{fig:phi_plots}b requires choosing the reference point ${\bf
  R}_j$ that sets the origin of a molecular coordinate system. The
dipole ${\bm\mu}_j$ is not sensitive to this choice, but its
contribution $\langle V_{\rm near}^{\rm dip}\rangle_q$ is. At higher
orders, both the multipole moment (e.g., ${\mathbf{K}}_j$) and its
contribution to the electric potential (e.g., $\langle V_{\rm
  near}^{\rm quad}\rangle_q$) depend on the choice of ${\bf R}_j$. For
$\langle V_{\rm near}^{\rm dip}\rangle_q$ and $\langle V_{\rm
  near}^{\rm quad}\rangle_q$ we find only a weak sensitivity for
reasonable choices of ${\bf R}_j$, i.e., points within the solute's
excluded volume that lie along the line of symmetry bisecting the
hydrogen atoms and running through the O atom.  Throughout this work
we will adopt the molecule's center of charge $\mbf{R}_j^{\rm (c)}
\equiv \sum_\alpha |q_\alpha|\mbf{r}_{j\alpha}/\sum_\alpha
|q_\alpha|$, which is displaced $\sim 0.3$\,\AA{} from the O atom, as
the reference point, which is shown schematically in
Fig.~\ref{fig:water_frame}a.  Fig.~\ref{fig:water_frame}b provides a
visual argument for this choice of molecular reference
frame. Isosurfaces are shown for the corresponding electric potential
\begin{equation}
  \phi^{\rm (2)}_j({\bf r}) =
  -{\bm\mu}_j\cdot\nabla\frac{1}{|{\bf r}-{\bf R}_j|}
  + \frac{1}{2}{\mathbf{K}}_j:\nabla\nabla\frac{1}{|{\bf r}-{\bf R}_j|},
\label{eqn:phi2}
\end{equation}
i.e., the multipole expansion of the potential generated by molecule
$j$, truncated at second order. The strong resemblance to water's
intramolecular geometry suggests that this low-order expansion
captures aspects of charge asymmetry essential to ion-specific
solvation. By contrast, an analogous second-order expansion with ${\bf
  R}_j$ set at the O atom gives rise to a potential that resembles
that of water much less closely (see SI).

These results encourage amending DCT to account for fluctuations in
local quadrupole density. The approach we describe below for doing so
is straightforward to generalize for higher-order multipoles. For the
sake of generality, we therefore introduce the $n^{\rm th}$-order
multipole moment of a water molecule as
\begin{widetext}
\begin{equation}
  \label{eqn:Ctensors}
  [{\bm\Gamma}^{(n)}]_{qrst\ldots} = \sum_{\alpha} q_\alpha
  [({\bf r}_{j\alpha}-{\bf R}_j)^n]_{qrst\ldots}
  + \delta_{qr} a_{st\ldots}
  + \delta_{qs} a'_{rt\ldots}
  + \delta_{rs} a''_{qt\ldots}
 + \ldots
\end{equation}
\end{widetext}
The constants $a$, $a'$, $a''$, etc. are all arbitrary, since the
moments ${\bm\Gamma}^{(n)}$ will always appear in contraction with
tensors formed from gradients of the electrostatic Green's
function. As with the quadrupole in Eq.~\ref{eqn:Qmol}, terms in
${\bm\Gamma}^{(n)}$ containing an identity in any two components
(e.g. $\delta_{qr}$) are therefore inconsequential. This contraction
also makes the ordering of indices in
$[{\bm\Gamma}^{(n)}]_{qrst\ldots}$ irrelevant.

\subsection{Elaborating DCT}
\label{subsec:ElabDCT}

Extending DCT to describe higher-order multipoles could be
accomplished most simply by adding a Gaussian quadrupole field ${\bf
  Q}_{\bf r}$.
A natural choice for the generalized Hamiltonian,
\begin{eqnarray}
  {\cal H}_{\rm dip+quad}[{\bf m_r},{\bf Q_r}] &=&  {\cal H}_{\rm dip}[{\bf m_r}] +
  \frac{1}{2\sigma_Q^2}\sum_{\bf r} {\bf Q}_{\bf r}:{\bf Q}_{\bf r}
  \nonumber\\ &-&
  \frac{1}{2}\sum_{\bf r} \sum_{{\bf r}'} {\bf Q}_{\bf r}:\bigg[
    \nabla\nabla\nabla \frac{1}{|{\bf r}-{\bf r}'|} \cdot{\bf m}_{{\bf r}'}
    \nonumber\\ &-&
    \frac{1}{4}\nabla\nabla\nabla\nabla \frac{1}{|{\bf r}-{\bf r}'|}
    : {\bf Q}_{{\bf r}'}
    \bigg]
  \label{equ:generalized-field-theory}
\end{eqnarray}
couples these fields through standard electrostatic interactions,
providing a bias that renormalizes dipolar linear response. Though
straightforward (and easily generalized to octupole density and
hexadecapole moments, etc.), this approach is unsatisfying in several
respects.

First, the electrostatic interaction between quadrupoles (and all
higher-order multipoles) diverges at short range in a way that is not
integrable in 3 dimensions, unlike the dipole-dipole interaction.
The field theory defined by Eq.~\ref{equ:generalized-field-theory}
would therefore require regularization, so that the
parameter $\sigma_Q$ sets a finite local quadrupole
susceptibility. Second, and more importantly, this theory preserves
the charge symmetry of standard DCT. Specifically, coupling to a
charged solute gives a total energy
$$
{\cal H}_{\rm dip+quad}
- \sum_{\bf r}\, {\bf E}_q({\bf r}) \cdot {\bf m_r} 
-\frac{1}{2} \sum_{\bf r}\, \nabla{\bf E}_q({\bf r}) : {\bf Q}_{\bf r}
$$
that is invariant to a simultaneous sign change of $q$, ${\bf m}_{\bf
  r}$, and ${\bf Q}_{\bf r}$. Just as for the Born model, ion
solvation energies would remain even in solute charge $q$.  Third, the
number of degrees of freedom proliferates in such a generalization as
higher order multipoles are included. At quadrupole order, the theory
involves 12 scalar variables at each point in space. Imposing expected
symmetries of ${\bf Q}_{\bf r}$ would reduce this number, but the fact
remains that adding detail (in the form of higher-order multipoles)
increases the theory's dimensionality. By contrast, an individual
water molecule, modeled as a rigid body, possesses only 3
non-translational degrees of freedom, regardless of how exhaustively
its electric potential $\phi({\bf r})$ is expanded in multipole
moments. These moments are not entirely independent variables; they
are instead tied together by molecular geometry. Such constraints
among molecular multipoles, we argue, are key to capturing charge
asymmetry at a field theoretic level.

\section{Developing a charge asymmetric field theory}
\label{sec:DevelopingFieldTheory}

\subsection{Multipole constraints and dipolar response}
\label{subsec:ConstraintsDipResponse}

The relationships among water's molecular multipoles can be easily
understood. As an illustration, consider the dipole and quadrupole of
an SPC/E water molecule. These moments are simply expressed in the
coordinate system of Fig.~\ref{fig:water_frame}a, ${\bm\mu} =
\mu\hat{\bm\mu}$ and ${\mathbf{K}} = K\hat{\bf x}\hat{\bf x}$, where
$\mu$ is the magnitude of the dipole vector, and $K$ is a scalar
constant. The unit vectors $\hat{\bm\mu}$ and $\hat{\bf x}$ point
parallel and perpendicular, respectively, to the line of symmetry
bisecting the hydrogen atoms. In the course of free molecular
rotation, $\hat{\bm\mu}$ and $\hat{\bf x}$ can both explore the entire
unit sphere, setting the range of possible realizations of ${\bm\mu}$
and ${\mathbf{K}}$. But if $\hat{\bm\mu}$ is fixed, $\hat{\bf x}$ can
explore only a unit circle orthogonal to $\hat{\bm\mu}$, limiting the
range of the tensor $\hat{\bf x}\hat{\bf x}$.  Quadrupole fluctuations
are thus partially constrained by the dipole's orientation, as are all
higher-order moments.

We imagine that multipole density fields like ${\bf m_r}$ and ${\bf
  Q_r}$ represent a coarse-grained view on a material's molecular
configuration. The coarse-graining procedure translates the constraint
detailed above between each molecule's dipole ${\bm\mu}_j$ and its
quadrupole ${\mathbf{K}}_j$ into a relationship between the fields
${\bf m_r}$ and ${\bf Q_r}$ -- a connection that is less strict and
more subtle than that between ${\bm\mu}_j$ and ${\mathbf{K}}_j$. We do
not attempt here to detail this connection between coarse-grained
fields.  Instead, we focus on the molecular constraint's influence on
the dipolar response of a single molecule. The result of this
molecular calculation will then be used to motivate a modification of
DCT.

\subsubsection{Dipole statistics of an isolated molecule}
\label{subsec:DipStatsIsolated}

Consider a single water molecule, at equilibrium, in an electric
potential $\phi_{\rm ext}({\bf r})$ that is generated by external charges.
Taking the molecule's center of charge to be fixed at a position ${\bf
  R}_0$, a multipole expansion expresses its energy as
\begin{equation}
h(\hat{\bm\mu},\hat{\bf x}) = \sum_{n=1}^\infty\frac{1}{n!}
\sum_{r,s,t,\ldots}[{\bm\Gamma}^{(n)}]_{rst\ldots}
\frac{\partial}{\partial x_r} \frac{\partial}{\partial x_s}
\frac{\partial}{\partial x_t}\ldots \phi_{\rm ext}({\bf
  r})\bigg|_{{\bf R}_0},
\label{equ:molecule-in-a-field}
\end{equation}
where the $n^{\rm th}$ term of the expansion involves $n$ spatial
derivatives, indexed by $r, s, t, \ldots$. We have in mind a model
with rigid intramolecular geometry, so that the unit vectors
$\hat{\bm\mu}$ and $\hat{\bf x}$ specify the entire set of multipole
moments ${\bm\Gamma}^{(n)}(\hat{\bm\mu},\hat{\bf x})$. We aim here to
integrate over one intramolecular degree of freedom ($\hat{\bf x}$),
while holding the other ($\hat{\bm\mu}$) fixed, to obtain a free
energy
\begin{equation}
h_{\rm eff}(\hat{\bm\mu}) = -k_{\rm B}T\ln
\int d\hat{\bf x}\,e^{-\beta h(\hat{\bm\mu},\hat{\bf x})}
\label{equ:effective-energy}
\end{equation}
that depends only on $\hat{\bm\mu}$. In doing so, we determine an
effective energy $h_{\rm eff}(\hat{\bm\mu})$ for the molecular dipole
in which fluctuations of all higher-order multipole moments have been
taken into account, along with the constraints that relate them.

The integration in Eq.~\ref{equ:effective-energy} is analytically
intractable, even for this simplified single-molecule scenario.  The
complicated electric field fluctuations generated by a liquid
environment at microscopic scales do not invite greatly simplifying
approximations. We nevertheless introduce two such assumptions, which
will allow us to capture the lowest-order influence of quadrupole (or
octupole, etc.) fluctuations on the statistics of molecular
dipoles. We first make a weak-field approximation,
\begin{widetext}
\begin{equation}
h_{\rm eff} \approx \langle h\rangle^{\rm unbiased}_{\hat{\bm\mu}}=
\sum_{n=1}^\infty \frac{1}{n!}\sum_{r,s,t,\ldots}
    [\langle{\bm\Gamma}^{(n)}\rangle^{\rm unbiased}_{\hat{\bm\mu}}]_{rst\ldots}
\frac{\partial}{\partial x_r}
\frac{\partial}{\partial x_s}
\frac{\partial}{\partial x_t}\ldots
\phi_{\rm ext}({\bf r})\bigg|_{{\bf R}_0},
\label{equ:h-eff}
\end{equation}
\end{widetext}
where $\langle \cdot \rangle^{\rm unbiased}_{\hat{\bm\mu}}$ denotes an
unbiased orientational average over $\hat{\bf x}$ subject to the
constraint of fixing $\hat{\bm\mu}$. We further assume that the
potential $\phi_{\rm ext}({\bf r})$ is slowly varying, so that the sum
over $n$ can be truncated at low order.

The symmetry of a water molecule causes many elements of
${\bm\Gamma}^{(n)}$ to vanish, regardless of the specific model
considered.  The irrelevance of isotropic contributions (e.g., terms
in $\mathbf{K}$ that are proportional to ${\bf I}$) causes many other
multipole elements to be unimportant. Furthermore, the contraction in
Eq.~\ref{equ:molecule-in-a-field} allows the indices of
$[{\bm\Gamma}^{(n)}]_{rst\ldots}$ to be permuted arbitrarily. As a
result, the class of elements relevant to $h$ at a given multipole
order is not large. Nontrivial contributions to ${\bm\Gamma}^{(n)}$
can all be written in terms of dyadic products involving an even
number $2m$ of factors $\hat{\bf x}$ together with $n-2m$ factors of
$\hat{\bm\mu}$ (e.g., $\hat{\bm\mu}\hat{\bm\mu}$ and $\hat{\bf
  x}\hat{\bf x}$ at order $n=2$,
$\hat{\bm\mu}\hat{\bm\mu}\hat{\bm\mu}$ and $\hat{\bm\mu}\hat{\bf
  x}\hat{\bf x}$ at order $n=3$, etc.). Relevant contributions to the
{\em average} moments $\langle{\bm\Gamma}^{(n)}\rangle^{\rm
  unbiased}_{\hat{\bm\mu}}$ in Eq.~\ref{equ:h-eff} then follow from
results of straightforward angular integration:
\begin{align}
  &\langle \hat{\bf x}\hat{\bf x}  \rangle^{\rm unbiased}_{\hat{\bm\mu}} =
  \frac{1}{2}({\bf I} - \hat{\bm\mu}\hat{\bm\mu}),
  \nonumber \\
  &\langle \hat{\bf x}\hat{\bf x} \hat{\bf x}\hat{\bf x}
  \rangle^{\rm unbiased}_{\hat{\bm\mu}} =
  \frac{3}{8}({\bf I} - \hat{\bm\mu}\hat{\bm\mu})
  ({\bf I} - \hat{\bm\mu}\hat{\bm\mu}), \qquad \ldots
\end{align}
where we have exploited the arbitrariness of index ordering in
$[\langle{\bm\Gamma}^{(n)}\rangle^{\rm
    unbiased}_{\hat{\bm\mu}}]_{rst\ldots}$.  Removing isotropic
contributions that vanish when contracted with gradients of $\phi_{\rm
  ext}({\bf r})$, we finally obtain an effective dipolar energy
\begin{equation}
  h_{\rm eff}
  \approx \sum_{n=1}^{\infty} b_n
  ({\bm\mu}\cdot\nabla)^n
 \phi_{\rm ext}({\bf r})\bigg|_{{\bf R}_0}
\label{heff-result}
\end{equation}
The form of this result is general for any SPC model that has the
symmetry of a water molecule. The values of coefficients $b_n$, on the
other hand, are model-specific; they are also sensitive to the choice
of reference point ${\bf R}_0$ defining the multipole expansion. For
the case of SPC/E water and ${\bf R}_0$ set at the molecule's center
of charge, $b_1=1$ and $b_2 \approx -0.6e^{-1}$.

In constructing a field theory in the next section, we will focus on a
truncation of the sum in Eq.~\ref{heff-result} at $n=2$,
\begin{equation}
  h_{\rm eff}^{(2)}({\bm\mu}) =
  {\bm\mu}\cdot\nabla \phi_{\rm ext} +
  b_2 ({\bm\mu}\cdot\nabla)^2
  \phi_{\rm ext}
\label{heff-quad}
\end{equation}
The first term in $h_{\rm eff}^{(2)}$ describes direct electrostatic
coupling between the molecular dipole and an electric field external
to that molecule. Its coarse-grained analog, appearing explicitly in
Eq.~\ref{eqn:dct-mic-int} and implicitly in Eq.~\ref{eqn:dct-mic}
through the dipole-dipole interaction, defines the nonlocal
interaction energy in DCT.  Correspondingly, this contribution is
charge symmetric -- a potential $\phi_{\rm ext} = q/r$ due to an
external point charge yields an energy that is invariant to inverting
the signs of both $q$ and ${\bm\mu}$.

The second term in $h_{\rm eff}^{(2)}$, by contrast, breaks charge
symmetry -- it effects response to an external point charge that is
not equivalent for $q>0$ and $q<0$. This nonlinear contribution
originates in fluctuations of $\hat{\bf x}$, which dictates the
molecular quadrupole. By integrating out quadrupole fluctuations,
we have thus obtained an effective dipolar energy that reflects the
asymmetric charge distribution within a water molecule.

Carrying out the summation in Eq.~\ref{heff-result} to higher order
generates a series of charge symmetric ($n$ odd) and charge
antisymmetric ($n$ even) terms. If a particular model and choice of
${\bf R}_0$ gives $b_2 = 0$ (as is the case for SPC/E water if one
chooses the oxygen atom as the reference point), charge asymmetry
would emerge first at hexadecapole order ($n=4$). If a particular
model features a completely charge-symmetric intramolecular geometry
(e.g., BNS water \cite{stillinger1974improved}) and ${\bf R}_0$ is set
at the center of charge, then $b_n = 0$ for all even values of $n$, so
that $h_{\rm eff}$ is appropriately equivalent for cation and anion
response.

\subsection{A constraint-inspired field theory}
\label{subsec:AsymDipTheory}

The analysis of single-molecule response we have presented suggests
important considerations for generalizing DCT.  Foremost, it indicates
that the introduction of quadrupolar fields (or other higher-order
multipole moments), as in Eq.~\ref{equ:generalized-field-theory},
should be accompanied by consideration of constraints dictated by
molecular geometry. The nature of these constraints is clear at the
molecular level, but an appropriate expression in terms of
coarse-grained fields like ${\bf m}_{\bf r}$ and ${\bf Q}_{\bf r}$ is
not obvious. If one were to impose strict constraints, such as ${\bf
  m}_{\bf r} \cdot {\bf Q}_{\bf r} = 0$ at each position ${\bf r}$,
then partition functions and response functions could be formulated
from Eq.~\ref{equ:generalized-field-theory} using methods that have
proven effective in other contexts
\cite{Chandler1993sjc,song1996gaussian}.  This approach would be
analytically challenging, however, since the nonlinear constraints we
have described prevent mapping onto a Gaussian theory simply by
introducing auxiliary fields as in
Refs.~\onlinecite{Chandler1993sjc,song1996gaussian}.

We will follow a different approach. Rather than taking the
constraints themselves from a molecular model, we instead take the
effective dipolar energy (Eq.~\ref{heff-quad}) they imply when local
quadrupole fluctuations are integrated out.  In doing so, we neglect
correlated fluctuations in the quadrupole field, in effect treating
${\bf Q}_{\bf r}$ and ${\bf Q}_{{\bf r}'}$ (with ${\bf r} \neq {\bf
  r}'$) as independent variables for a given realization of the dipole
field. Focusing in this way on quadrupolar response to ${\bf m}_{\bf
  r}$ alone conforms to the spirit of the multipole expansion on which
our perspective is based. The resulting charge-asymmetric,
field-theoretic Hamiltonian follows from Eq.~\ref{heff-quad},
\begin{equation}
  {\cal H}[{\bf m}_{\bf r}] =
  {\cal H}_{\rm dip}[{\bf m}_{\bf r}] - \sum_{\bf r}\, {\bf E}_q({\bf r}) \cdot {\bf m}_{\bf r}
  - b \sum_{\bf r} \, 
  {\bf m}_{\bf r} \, {\bf m}_{\bf r} : \nabla {\bf E}({\bf r})
  \label{equ:full-field-theory}
\end{equation}
where ${\bf E} = {\bf E}_q + {\bf E}_{\rm dip}$ is the total electric
field at ${\bf r}$, including contributions from the solute and from
the dipole field,
$$
{\bf E}_{\rm dip}({\bf r}) = -\sum_{{\bf r}'} \nabla \nabla'
\frac{1}{|{\bf r}-{\bf r}'|} \cdot {\bf m}_{\bf r}'
$$
We will not attempt here to derive or motivate a value for the
parameter $b$, whose connection to the molecular parameter $b_2$ is
conceptually but not quantitatively clear.

Eq.~\ref{equ:full-field-theory} is the central result of this
paper. It defines a field theory that is charge asymmetric in accord
with the asymmetric response of an isolated water molecule. It
respects the rotational symmetry of the liquid state and is simple to
express, but
analysis is made unwieldy by the final sum in
Eq.~\ref{equ:full-field-theory}, which features
coupling of the external field to a bilinear functional
of the dipole field as well as cubic, spatially nonlocal interactions
among the field variables.
To make exploratory progress, we introduce two additional
approximations.  First, we replace the fluctuating total electric
field in Eq.~\ref{equ:full-field-theory} with a constant, screened
external field, ${\bf E}({\bf r}) \approx {\bf E}_q/\epsilon$, that
would result on average from linear dielectric response. This
replacement removes a nonlinearity of third order in the dipole field,
while preserving nonlinear {\em response} to the solute's charge. It
also limits the complications we have added to a spatially local
functional of the field ${\bf m}_{\bf r}$.

With this simplification, the Hamiltonian in
Eq.~\ref{equ:full-field-theory} becomes bilinear in the dipole field,
whose statistics are therefore Gaussian. Analysis remains challenging,
however, because the effective dipolar coupling generates localized
normal modes that are not easily anticipated.  The fluctuation
spectrum of ${\bf m}_{\bf r}$ thus changes as ${\bf E}_q$ is
introduced, producing a complicated nonlinear response that breaks
charge symmetry.  We simplify further by taking a variational
approach, introducing a more tractable reference system
\begin{equation}
  {\cal H}_{\rm ref}[{\bf m_r}] = {\cal H}_{\rm dip}[{\bf m_r}] + \tilde{q} \sum_{\bf r}\, {\bf m}_{\bf
    r} \cdot \nabla r^{-1}
\end{equation}
${\cal H}_{\rm ref}$ describes the response of a conventional
dielectric continuum to a solute with {\em effective} charge
$\tilde{q}$.

We determine an optimal choice of the variational
parameter $\tilde{q}$ from the Gibbs-Bogoliubov bound,
$$
\ln\mathcal{Z} \ge \ln\mathcal{Z}_{\rm ref} -
\beta\langle\Delta\mathcal{H}\rangle_{\rm ref}
$$
on the partition function $\mathcal{Z}$ and its counterpart
$\mathcal{Z}_{\rm ref}$ for the reference system. Here,
$\Delta\mathcal{H} = \mathcal{H} - \mathcal{H}_{\rm ref}$, and
$\langle\cdot\rangle_{\rm ref}$ denotes an ensemble average in the
reference system.  Evaluating $\langle\Delta\mathcal{H}\rangle_{\rm
  ref}$ requires calculating, and appropriately summing, both $\langle
{\bf m}_{\bf r} \rangle_{\rm ref}$ and $\langle {\bf m}_{\bf r} {\bf
  m}_{\bf r} \rangle_{\rm ref}$.  The former, $\langle {\bf m}_{\bf r}
\rangle_{\rm ref} = {\bf E}_q(\epsilon-1)/(4\pi\epsilon)$ is simple to
compute and manipulate, both on- and off-lattice. The latter involves
the response function $\chi({\bf r},{\bf r}') = \langle \delta{\bf
  m}_{\bf r} \delta{\bf m}_{{\bf r}'} \rangle_{\rm ref}$, where
$\delta{\bf m}_{\bf r} = {\bf m}_{\bf r} - \langle {\bf m}_{\bf r}
\rangle_{\rm ref}$. In the presence of a volume-excluding solute,
$\chi({\bf r},{\bf r}')$ is generally complicated, and in the
off-lattice case it is singular for ${\bf r} = {\bf r}'$. But with
space treated discretely it can be written compactly for a solute that
occupies a single lattice cell. Placing this solute at the origin, we
have \cite{song1996gaussian}
$$
\chi({\bf r},{\bf r}') = - \frac{3 v^3}{\beta\epsilon (2\epsilon + 1)}
\bigg( \frac{\epsilon-1}{4\pi}\bigg)^3
\nabla \nabla \frac{1}{r} \cdot \nabla \nabla \frac{1}{r}
$$
for ${\bf r}\neq 0$. Approximating sums
$\sum_{{\bf r}\neq 0}$ as integrals $v^{-1} \int_{r<R} d{\bf r}$,
we obtain
$$
\tilde{q} = \frac{q}{1+B q}
$$
with $B = -b (v/R^3) (\epsilon-1)/(4\pi\epsilon)$, and
\begin{align}
F_{\rm chg}^{\rm (var)}(q) &\approx F_{\rm chg}^{\rm (ref)}(q) +
\langle\Delta\mathcal{H}\rangle_{\rm ref} \nonumber \\
&=
-\frac{q\tilde{q}}{2R}\frac{\epsilon-1}{\epsilon}
- \frac{3b q k_{\rm B}T}{2\pi^2\epsilon}
\frac{(\epsilon-1)^3}{2\epsilon+1}.
\label{eqn:Fvar}
\end{align}
For simplicity we have taken $R = v^{1/3}/2$. Reasonable alternatives,
such as $R = (3v/4\pi)^{1/3}$ yield similar results.

Eq.~\ref{eqn:Fvar} includes a term linear in $q$, whose coefficient
could be regarded as a contribution to the neutral cavity potential
$\langle V\rangle_0$.  Since we have made no attempt to include
contributions from distant interfaces, this term cannot offer a full
accounting of charge asymmetry in the limit $q\rightarrow 0$. In the
same spirit as the modified Born model in Eq.~\ref{eqn:born-qV0}, we
could replace it with the correct neutral cavity potential,
\begin{equation}
  \label{eqn:Fvar-qV0}
  F_{\rm chg}^{\rm (var)}(q; \langle V\rangle_0) \approx -\frac{q\tilde{q}}{2R}\frac{\epsilon-1}{\epsilon} + q\langle V\rangle_0.
\end{equation}
This modification is only significant at very small values of $q$. Predictions
for fully charged ions ($q=\pm e$) are essentially unaffected.

\section{Numerical results}
\label{sec:Assessing}

Predictions of the variational result in Eq.~\ref{eqn:Fvar-qV0} depend
on input parameters $R$ and $b$, which set the effective solute size
and the strength of nonlinearity due to quadrupole fluctuations. We
will treat these parameters as we did the dielectric radius $R$ of the
Born model in Sec.~\ref{subsec:AsymSolv}. Specifically, we require
consistency across ions with a given volume-excluding radius $R_0$ but
otherwise adjust $R$ and $b$ to obtain the best possible agreement
with results from molecular simulation.

\begin{figure}[!tb]
  \includegraphics[width=8cm]{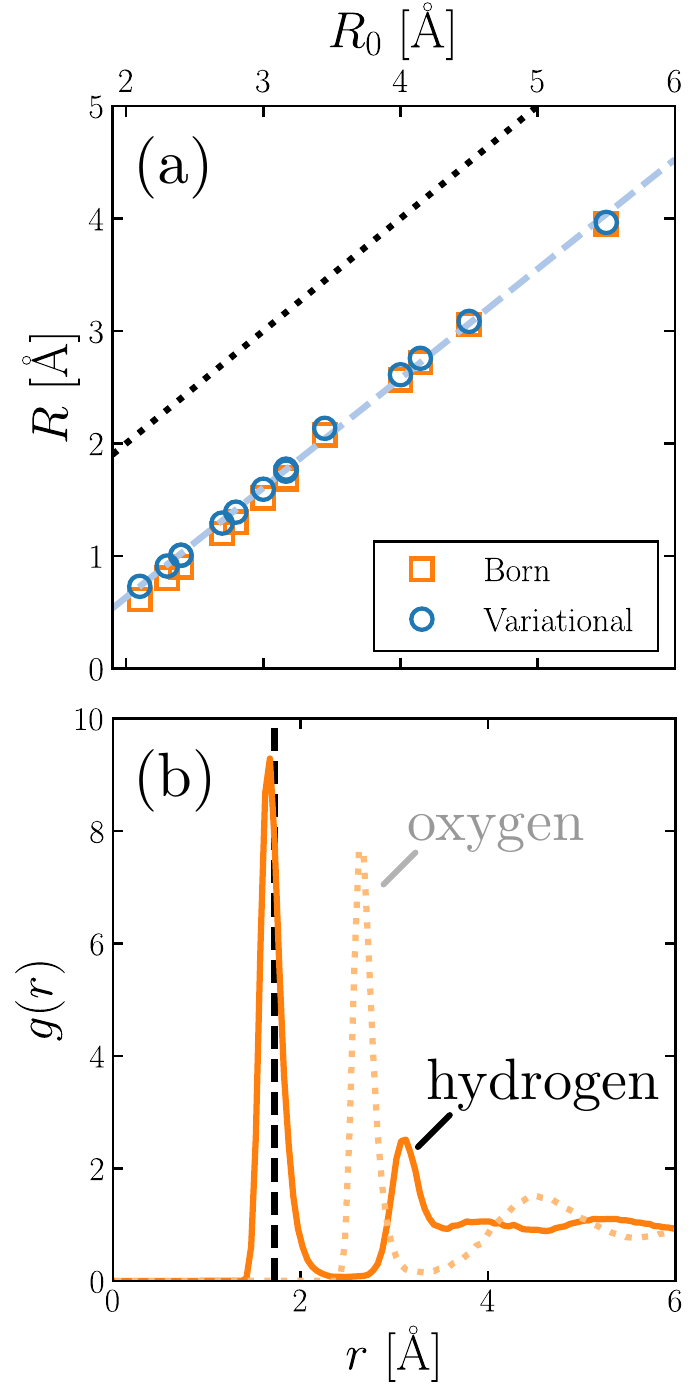}
  \caption{The optimal dielectric radius $R$ is generally smaller than
    the radius $R_0$ of a solute's excluded volume.  (a) The
    relationship between $R$ and $R_0$, determined by fitting
    theoretical results to computer simulations, is approximately
    linear. The dotted line indicates $R=R_0$, and the dashed line
    shows $R\approx 0.97R_0 - 1.31$\,\AA.  Results are shown for the
    Born model (Eq.~\ref{eqn:born-qV0}) and for our variational theory
    (Eq.~\ref{eqn:Fvar-qV0}).  (b) The solute-solvent radial
    distribution function [$g(r)$] suggests $R$ roughly corresponds to
    the distance of closest approach of water's hydrogen atoms to the
    solute. Solid and dotted lines show solute-hydrogen $g(r)$ and
    solute-oxygen $g(r)$, respectively, for $R_0=3.17$\,\AA{} and
    $q=-e$. The vertical dashed line indicates the best-fit value of
    $R$ for this solute size.}
  \label{fig:RDFs}
\end{figure}

For any physically well-founded theory, we expect the optimal choice
of dielectric radius $R$ to be similar, but not necessarily identical,
to the radius $R_0$ of molecular volume exclusion. Indeed,
Fig.~\ref{fig:RDFs}a shows that $R$ for our nonlinear variational
theory depends on $R_0$ in almost precisely the same way as for the
Born model. Across the range of solute sizes considered, we find that
dielectric and volume-excluding radii differ by a nearly constant
offset, $R \approx R_0 - 1.31$\,\AA.
\footnote{We again equate $R_0$ with the Lennard-Jones diameter
    for ion-water interactions. An alternative estimate of $R_0$ from
    the first peak of ion-oxygen radial distribution functions yields
    a similar relationship between $R$ and $R_0$, but with a smaller
    offset.}
With this offset, $R$ corresponds
suggestively to the distance of closest approach between the solute
and the hydrogen atoms of surrounding water molecules, as shown in
Fig.~\ref{fig:RDFs}b for $q = -e$ and $R_0 = 3.17$\,\AA{} (examples
for different choices of $R_0$ are given in the SI). The idea that an
optimal dielectric radius may appear smaller than $R_0$ owing to the
longer reach of water's hydrogen atoms has been proposed and discussed
before \cite{latimer1939free,rajamani2004size,bardhan2012affine}, but
for the specific case of negatively charged solutes, whose solvation
favors molecular orientations that place hydrogen atoms as close to
the solute as possible. In our case, we stress that the same
dielectric radius is used for cations and anions that have the same
volume-excluding size. Charge asymmetry is an emergent, rather than
engineered, feature of this approach.

\begin{figure}[!tb]
  \includegraphics[width=8cm]{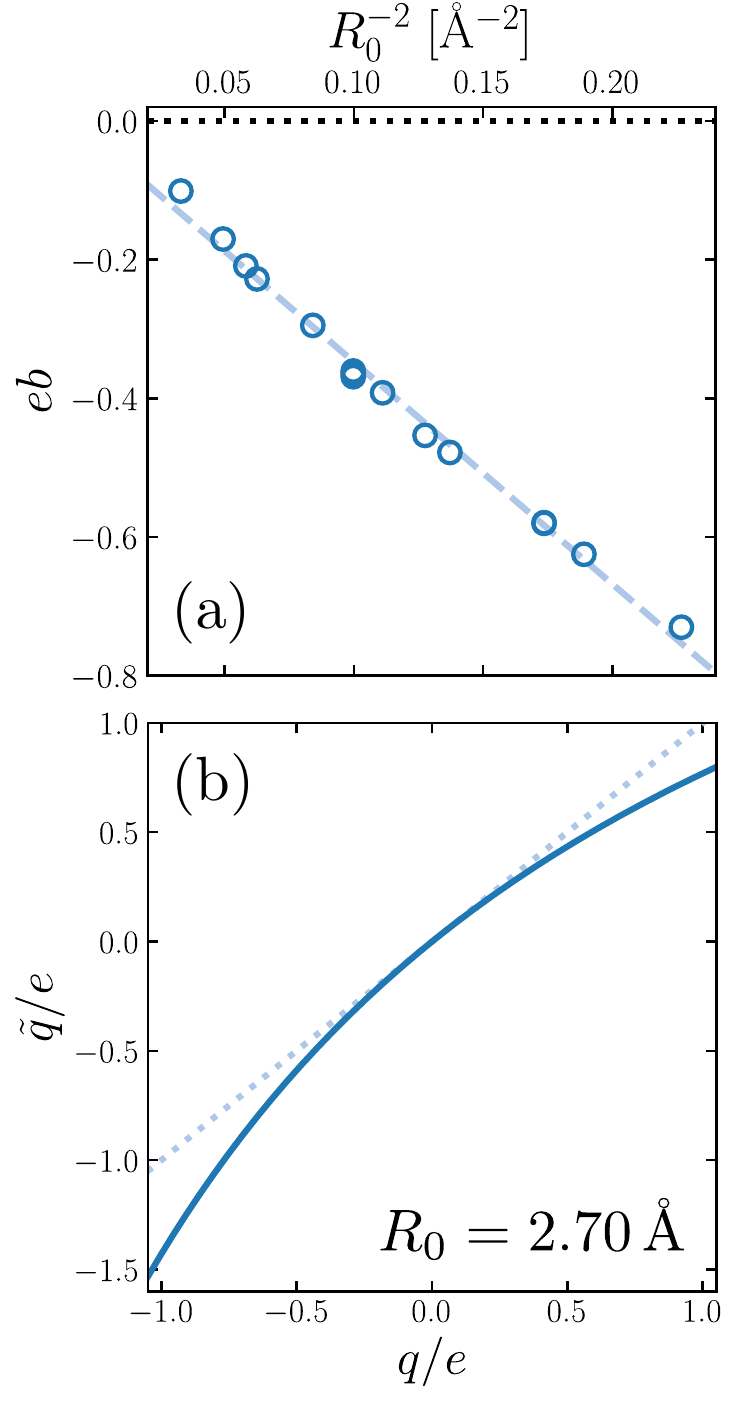}
  \caption{(a) The nonlinearity parameter $b$ (see
    Eq.~\ref{equ:full-field-theory}), determined by fitting
    theoretical results to computer simulations, varies linearly with
    $R_0^{-2}$. The dashed line shows $eb=-0.03-3.20R_0^{-2}$. (b) The
    resulting effective variational charge $\tilde{q}$ is greater in
    magnitude for anions than it is for cations, as shown for a solute
    with volume excluding radius $R_0=2.70$\,\AA{} (solid line). The
    dotted line indicates $\tilde{q}=q$.}
  \label{fig:b-vs-R02inv}
\end{figure}

Because the optimal choice of $R$ aligns closely with that of the Born
model, we view the nonlinear theory of Eq.~\ref{equ:full-field-theory}
as adding a single adjustable parameter, namely $b$. We anticipate
that $b$, which has units of inverse charge, should be roughly of
order $1/e$.  We also expect that $b$ should decay in magnitude as
solute size $R_0$ increases, both because near-field contributions are
less prominent for large ions and because linear response theory is
already successful in this limit. The origin of $b$ in constraints of
molecular geometry, which are not at all transparent at a
field-theoretic level, makes it difficult to develop further {\em a
  priori} expectations.  Maximizing agreement of
Eq.~\ref{eqn:Fvar-qV0} with simulation results for $F_{\rm chg}(q)$,
we find empirically that $b \approx e^{-1} (-0.03 - 3.20 R_0^{-2})$ to
a very good approximation, as shown in Fig.~\ref{fig:b-vs-R02inv}a. As
a practical matter, this simple and quantitatively successful fit
allows accurate application of the variational result in
Eq.~\ref{eqn:Fvar-qV0} to arbitrary $R_0$ without any further
fitting. Physically, the observed scaling of $b$ with $R_0$ is
intriguing, but we cannot offer a compelling explanation.

\begin{figure}[tb]
  \includegraphics[width=8cm]{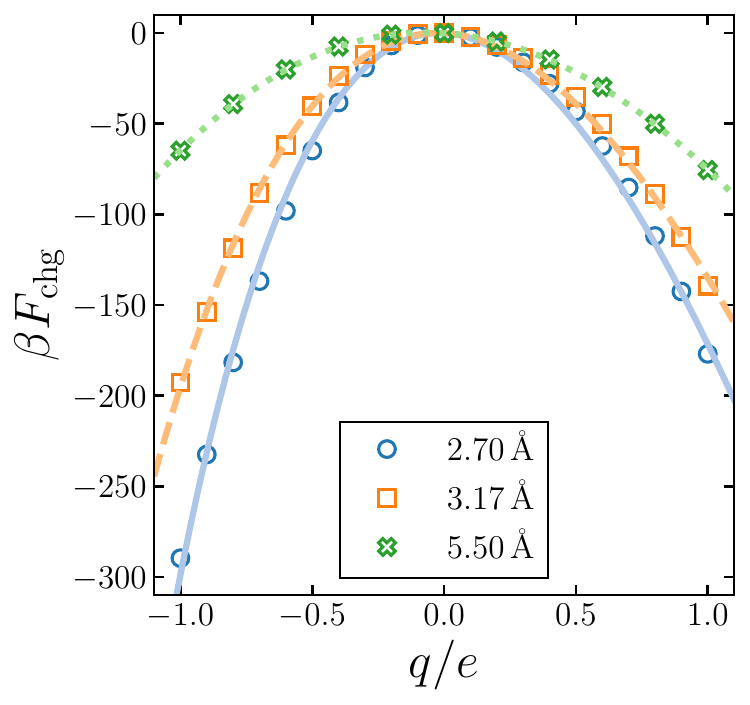}  
  \caption{Solute charging free energy $F_{\rm chg}^{\rm (var)}$
    estimated from the approximate variational solution to our
    nonlinear field theory, for different solute sizes $R_0$ as
    indicated in the legend. Symbols show results from
    simulations. Lines indicate best fits of $F_{\rm chg}^{\rm (var)}$
    (Eq.~\ref{eqn:Fvar-qV0}) to the simulation results. For smaller
    solute sizes, $F_{\rm chg}^{\rm (var)}$ significantly improves
    upon $F_{\rm chg}^{\rm (Born)}$, especially for $q=\pm e$ (see
    Fig.~\ref{fig:fe_born} and SI).}
  \label{fig:Fvar}
\end{figure}

With these fitted values of $R$ and $b$, the effective variational
charge $\tilde{q}$ is larger in magnitude for $q=-e$ than for $q=e$,
as shown in Fig.~\ref{fig:b-vs-R02inv}b for a solute with volume
excluding radius $R_0 = 2.70$\,\AA.  The resulting charge
asymmetry therefore favors solvation of fully charged anions over
cations with the same volume-excluding size, as observed in computer
simulations. Fig.~\ref{fig:Fvar} shows a detailed comparison of
charging free energies obtained from simulation and from the nonlinear
variational theory. For all solute sizes considered, and across the
entire range $q=-e$ to $q=e$, the agreement is excellent.  For the
largest solute, $R_0 = 5.5$\,\AA, there is little room for improvement
over the linear response prediction Eq.~\ref{eqn:born-qV0}; a small
but noticeable charge-asymmetric response in simulation results is
nonetheless captured well by our variational result.  For the smaller
solutes, $R_0 = 2.7$\,\AA{} and $R_0 = 3.17$\,\AA, improvement over
the Born model is dramatic. Discrepancies between simulation and the
nonlinear field theory result certainly remain, but the qualitative
shortcomings of DCT have essentially been erased.

\section{Discussion and Outlook}
\label{sec:Outlook}

Our aim in this article has been to address a key failing of DCT -- a
fundamental lack of charge asymmetry in ion solvation -- while
preserving its conceptual simplicity. Computer simulations indicate
that this asymmetry originates in induced polarization of the solvent
which is not simply a linear functional of the electric field exerted
by a charged solute.  Motivated by the influence of quadrupole
fluctuations and constraints of molecular geometry on statistics of
the solvent dipole field, the effective Hamiltonian presented in
Eq.~\ref{equ:full-field-theory} adds the kind of sensitivity to such
nonlinear response that is required to capture charge-asymmetric
solvation. Our approximate solution to this model, obtained by a
variational procedure, gives a charging free energy
(Eq.~\ref{eqn:Fvar-qV0}) with the same basic form as the standard Born
model (Eq.~\ref{eqn:born-qV0}), but with an effective ion charge that
is renormalized by nonlinear response. Setting the strength $b$ of the
nonlinearity to be a simple function of ion size, we obtain close
quantitative agreement with results of computer simulations.

While we believe our approach is original, it is certainly not the
only way to achieve charge asymmetric solvation energies. In many
previous efforts, asymmetry was introduced by hand.  Latimer, Pitzer
and Slansky\cite{latimer1939free} amended the Born model by assigning
different dielectric radii for anions and cations of the same size, an
approach that has been adopted in subsequent theoretical studies.  The
piecewise-affine response model\cite{bardhan2012affine} of Bardhan
\etal{} follows a similar spirit, empirically adjusting the nature of
electrostatic response as a solute's charge is varied.  We recently
demonstrated that an analogous treatment of interfacial solvation
performs reasonably well in describing ion-specific adsorption to the
air-water surface \cite{cox2020assessing}.  Also inspired by the
constraints between water's molecular multipoles, Mukhopadhyay \etal{}
introduced charge asymmetry into both the Born
\cite{MukhopadhyayOnufriev2012sjc} and generalized Born
\cite{mukhopadhyay2014introducing} models via a scaling factor that
depends upon the sign of the solute's charge. Similar to our approach,
the dielectric radius is also independent of the solute's charge. In
all these approaches, however, charge asymmetry was built in \emph{a
  posteriori}, whereas it is an emergent feature of the model defined
by Eq.~\ref{equ:full-field-theory}.

Fluctuations in a solvent's polarization and in its density are both
advanced at microscopic scales by rearrangement of discrete molecular
structures; they are therefore tied together intimately. In this paper
we have taken an electrostatic perspective on the nonlinear response
to solute charging, in which polarization fluctuations are
renormalized by degrees of freedom that can be described in terms of
electrostatic multipoles. Polarization statistics can of course also
be complicated by the influence of microscopic density fluctuations,
as highlighted by the sensitivity of dielectric suspectibility to
volume exclusion \cite{song1996gaussian}. Work by Dinpajooh and
Matyushov \cite{dinpajooh2015free} emphasizes that these biases are
not completely distinct, suggesting the interesting possibility that
the quadrupole-mediated response we have analyzed might be conceived
alternatively in terms of microscopic density fluctuations. More
recently, Duignan and Zhao have found that the degree of charge
asymmetry in simple point charge models can be drastically reduced by
shifting the center for volume exclusion on the water molecule from
the oxygen atom toward the hydrogen atoms \cite{duignan2020born}. This
sensitivity is distinct from that discussed in
Sec.~\ref{subsec:DipStatsIsolated}, which arises from truncating the
sum in Eq.~\ref{heff-result} at second order. In principle, however, a
field theory that is insensitive to $\mbf{R}_0$ could be constructed
by including all higher order contributions, even if its analysis
becomes intractable.

More generally, the interplay between density and polarization
response generates a spectrum of solvation behaviors, ranging from
hydrophobic effects at one extreme to small ion solvation at the
other. A lack of theoretical methods and tractable models that
successfully span this range stands as a one of the most severe
challenges limiting computational biophysics and nanoscience.  While
research on hydrophobic effects remains
active\cite{pratt2016statistical}, field theoretic approaches to the
underlying density fluctuations have matured greatly in recent years
\cite{LumWeeks1999sjc,chandler2005interfaces,VarillyChandler2011sjc,VaikuntanathanGeissler2014sjc,VaikuntanathanGeissler2016sjc}. The
powerful tools they provide do not yet have counterparts in an
electrostatic context, a gap that our work seeks to help fill. While
much remains to be done in refining the nonlinear theory we have
formulated and in developing practical methods to solve it,
the work presented here is in our view a meaningful step towards
placing theories for electrostatic and hydrophobic solvation on
comparable footing. As such, it advances the development of
efficient computational techniques that apply across the entire
hydrophobic/hydrophilic spectrum. 

\section{Methods}
\label{sec:Methods}

All simulations used the SPC/E water model
\cite{BerendsenStraatsma1987sjc} and were performed with the LAMMPS
simulation package\cite{plimpton1995sjc}.  Simulations comprised 64,
256 or 512 water molecules plus a single solute, such that the total
number density was $\rho = 0.03333$\,\AA$^{-3}$. Our model solute is a
Lennard-Jones particle,
\begin{equation}
  \label{eqn:LJ}
  u(r) = 4\varepsilon[(R_0/r)^{12}-(R_0/r)^{6}],
\end{equation}
where $r$ is the distance between the center of the solute (where the
solute charge is also located) and the oxygen atom of the water
molecule. We set $\varepsilon = 0.1553$\,kcal/mol (the same as SPC/E
water) for all solutes investigated, but varied $R_0$ as indicated
throughout the manuscript. Full 3D periodic boundary conditions with
particle-particle particle-mesh Ewald summation was used throughout
\cite{HockneyEastwood1988sjc,kolafa1992cutoff}, with a homogeneous
background charge to neutralize the system. Simulations of 5\,ns in
length, with a periodic cell of side length $L$, were performed with
$q/e = -1.0, -0.9,\ldots,+0.9,+1.0$. (For $R_0\ge 4.5$\,\AA{} we used
$q/e = -1.0, -0.8,\ldots,+0.8,+1.0$). The charging free energy
$F^{(L)}_{\rm chg}$ was then computed using the MBAR algorithm
\cite{ShirtsChodera2008sjc}, as described previously in
Ref~\onlinecite{cox2018interfacial}. The quantity $F^{(L)}_{\rm chg}$
suffers from severe finite size effects. It has previously been shown
\cite{HummerGarcia1996sjc,HunenbergerMcCammon1999sjc,cox2018interfacial}
that the quantity,
\begin{equation}
  \label{eqn:Fmacro}
  F_{\rm chg} = F^{(L)}_{\rm chg} + \frac{q}{2}\dctfac\phi_{\rm wig} + qV_{\rm surf}
\end{equation}
accurately estimates the macroscopic limit $L\rightarrow\infty$,
including the effects of distant interfaces. In Eq.~\ref{eqn:Fmacro},
the Wigner potential $\phi_{\text{wig}}/q$ is defined as the
electrostatic potential at the site of a unit point charge due to all
of its periodic replicas and a homogeneous background charge that acts
to neutralize the primitive cell.  The surface potential $V_{\rm
  surf}=-590$\,mV \cite{remsing2014role} was determined by numerically
integrating the solvent's charge density profile
$\langle\rho_{\text{solv}}(z)\rangle$ according to $V_{\rm surf} =
4\pi \int_{z_{\rm vap}}^{z_{\rm liq}}\!\mrm{d}z\,
\langle\rho_{\text{solv}}(z)\rangle z$, where $z_{\rm liq}$ and
$z_{\rm vap}$ denote locations on either side of a neat liquid/vapor
interface. Nonlinear curve fitting to obtain optimal choices of $R$
and $b$ (see Eqs.~\ref{eqn:born-qV0} and~\ref{eqn:Fvar-qV0}) was
performed using the True Region Reflective algorithm
\cite{branch1999subspace}, as implemented in SciPy's `curve\_fit'
routine \cite{virtanen2020scipy}.

\begin{acknowledgments}
  S.J.C (02/15 to 09/17) and P.L.G were supported by the
  U.S. Department of Energy, Office of Basic Energy Sciences, through
  the Chemical Sciences Division (CSD) of Lawrence Berkeley National
  Laboratory (LBNL), under Contract DE-AC02-05CH11231. K.K.M is
  supported by Director, Office of Science, Office of Basic Energy
  Sciences, of the U.S. Department of Energy under contract
  No. DEAC02-05CH11231. From 10/17 to 02/21, S.J.C. was supported by a
  Royal Commission for the Exhibition of 1851 Research Fellowship.
\end{acknowledgments}

\section*{Data Availability Statement}

The data that supports the findings of this study are available within
the article and its supplementary material. Data for the charging free
energies and a Python analysis script are openly available at the
University of Cambridge Data Repository,
\url{https://doi.org/10.17863/CAM.66169}.

\bibliography{../ver3/cox}

\begin{thebibliography}{64}%
\makeatletter
\providecommand \@ifxundefined [1]{%
 \@ifx{#1\undefined}
}%
\providecommand \@ifnum [1]{%
 \ifnum #1\expandafter \@firstoftwo
 \else \expandafter \@secondoftwo
 \fi
}%
\providecommand \@ifx [1]{%
 \ifx #1\expandafter \@firstoftwo
 \else \expandafter \@secondoftwo
 \fi
}%
\providecommand \natexlab [1]{#1}%
\providecommand \enquote  [1]{``#1''}%
\providecommand \bibnamefont  [1]{#1}%
\providecommand \bibfnamefont [1]{#1}%
\providecommand \citenamefont [1]{#1}%
\providecommand \href@noop [0]{\@secondoftwo}%
\providecommand \href [0]{\begingroup \@sanitize@url \@href}%
\providecommand \@href[1]{\@@startlink{#1}\@@href}%
\providecommand \@@href[1]{\endgroup#1\@@endlink}%
\providecommand \@sanitize@url [0]{\catcode `\\12\catcode `\$12\catcode
  `\&12\catcode `\#12\catcode `\^12\catcode `\_12\catcode `\%12\relax}%
\providecommand \@@startlink[1]{}%
\providecommand \@@endlink[0]{}%
\providecommand \url  [0]{\begingroup\@sanitize@url \@url }%
\providecommand \@url [1]{\endgroup\@href {#1}{\urlprefix }}%
\providecommand \urlprefix  [0]{URL }%
\providecommand \Eprint [0]{\href }%
\providecommand \doibase [0]{http://dx.doi.org/}%
\providecommand \selectlanguage [0]{\@gobble}%
\providecommand \bibinfo  [0]{\@secondoftwo}%
\providecommand \bibfield  [0]{\@secondoftwo}%
\providecommand \translation [1]{[#1]}%
\providecommand \BibitemOpen [0]{}%
\providecommand \bibitemStop [0]{}%
\providecommand \bibitemNoStop [0]{.\EOS\space}%
\providecommand \EOS [0]{\spacefactor3000\relax}%
\providecommand \BibitemShut  [1]{\csname bibitem#1\endcsname}%
\let\auto@bib@innerbib\@empty
\bibitem [{\citenamefont {Tielrooij}\ \emph {et~al.}(2010)\citenamefont
  {Tielrooij}, \citenamefont {Garcia-Araez}, \citenamefont {Bonn},\ and\
  \citenamefont {Bakker}}]{tielrooij2010cooperativity}%
  \BibitemOpen
  \bibfield  {author} {\bibinfo {author} {\bibfnamefont {K.}~\bibnamefont
  {Tielrooij}}, \bibinfo {author} {\bibfnamefont {N.}~\bibnamefont
  {Garcia-Araez}}, \bibinfo {author} {\bibfnamefont {M.}~\bibnamefont {Bonn}},
  \ and\ \bibinfo {author} {\bibfnamefont {H.}~\bibnamefont {Bakker}},\
  }\href@noop {} {\bibfield  {journal} {\bibinfo  {journal} {Science}\ }\textbf
  {\bibinfo {volume} {328}},\ \bibinfo {pages} {1006} (\bibinfo {year}
  {2010})}\BibitemShut {NoStop}%
\bibitem [{\citenamefont {Otten}\ \emph {et~al.}(2012)\citenamefont {Otten},
  \citenamefont {Shaffer}, \citenamefont {Geissler},\ and\ \citenamefont
  {Saykally}}]{OttenSaykally2012sjc}%
  \BibitemOpen
  \bibfield  {author} {\bibinfo {author} {\bibfnamefont {D.~E.}\ \bibnamefont
  {Otten}}, \bibinfo {author} {\bibfnamefont {P.~R.}\ \bibnamefont {Shaffer}},
  \bibinfo {author} {\bibfnamefont {P.~L.}\ \bibnamefont {Geissler}}, \ and\
  \bibinfo {author} {\bibfnamefont {R.~J.}\ \bibnamefont {Saykally}},\
  }\href@noop {} {\bibfield  {journal} {\bibinfo  {journal} {Proc. Natl. Acad.
  Sci. USA}\ }\textbf {\bibinfo {volume} {109}},\ \bibinfo {pages} {701}
  (\bibinfo {year} {2012})}\BibitemShut {NoStop}%
\bibitem [{\citenamefont {Verreault}, \citenamefont {Hua},\ and\ \citenamefont
  {Allen}(2012)}]{verreault2012conventional}%
  \BibitemOpen
  \bibfield  {author} {\bibinfo {author} {\bibfnamefont {D.}~\bibnamefont
  {Verreault}}, \bibinfo {author} {\bibfnamefont {W.}~\bibnamefont {Hua}}, \
  and\ \bibinfo {author} {\bibfnamefont {H.~C.}\ \bibnamefont {Allen}},\
  }\href@noop {} {\bibfield  {journal} {\bibinfo  {journal} {J. Phys. Chem.
  Lett.}\ }\textbf {\bibinfo {volume} {3}},\ \bibinfo {pages} {3012} (\bibinfo
  {year} {2012})}\BibitemShut {NoStop}%
\bibitem [{\citenamefont {Piatkowski}\ \emph {et~al.}(2014)\citenamefont
  {Piatkowski}, \citenamefont {Zhang}, \citenamefont {Backus}, \citenamefont
  {Bakker},\ and\ \citenamefont {Bonn}}]{piatkowski2014extreme}%
  \BibitemOpen
  \bibfield  {author} {\bibinfo {author} {\bibfnamefont {L.}~\bibnamefont
  {Piatkowski}}, \bibinfo {author} {\bibfnamefont {Z.}~\bibnamefont {Zhang}},
  \bibinfo {author} {\bibfnamefont {E.~H.}\ \bibnamefont {Backus}}, \bibinfo
  {author} {\bibfnamefont {H.~J.}\ \bibnamefont {Bakker}}, \ and\ \bibinfo
  {author} {\bibfnamefont {M.}~\bibnamefont {Bonn}},\ }\href@noop {} {\bibfield
   {journal} {\bibinfo  {journal} {Nature Commun.}\ }\textbf {\bibinfo {volume}
  {5}},\ \bibinfo {pages} {4083} (\bibinfo {year} {2014})}\BibitemShut
  {NoStop}%
\bibitem [{\citenamefont {McCaffrey}\ \emph {et~al.}(2017)\citenamefont
  {McCaffrey}, \citenamefont {Nguyen}, \citenamefont {Cox}, \citenamefont
  {Weller}, \citenamefont {Alivisatos}, \citenamefont {Geissler},\ and\
  \citenamefont {Saykally}}]{mccaffrey2017mechanism}%
  \BibitemOpen
  \bibfield  {author} {\bibinfo {author} {\bibfnamefont {D.~L.}\ \bibnamefont
  {McCaffrey}}, \bibinfo {author} {\bibfnamefont {S.~C.}\ \bibnamefont
  {Nguyen}}, \bibinfo {author} {\bibfnamefont {S.~J.}\ \bibnamefont {Cox}},
  \bibinfo {author} {\bibfnamefont {H.}~\bibnamefont {Weller}}, \bibinfo
  {author} {\bibfnamefont {A.~P.}\ \bibnamefont {Alivisatos}}, \bibinfo
  {author} {\bibfnamefont {P.~L.}\ \bibnamefont {Geissler}}, \ and\ \bibinfo
  {author} {\bibfnamefont {R.~J.}\ \bibnamefont {Saykally}},\ }\href@noop {}
  {\bibfield  {journal} {\bibinfo  {journal} {Proc. Natl. Acad. Sci. USA}\
  }\textbf {\bibinfo {volume} {114}},\ \bibinfo {pages} {13369} (\bibinfo
  {year} {2017})}\BibitemShut {NoStop}%
\bibitem [{\citenamefont {Chen}\ \emph {et~al.}(2016)\citenamefont {Chen},
  \citenamefont {Okur}, \citenamefont {Gomopoulos}, \citenamefont
  {Macias-Romero}, \citenamefont {Cremer}, \citenamefont {Petersen},
  \citenamefont {Tocci}, \citenamefont {Wilkins}, \citenamefont {Liang},
  \citenamefont {Ceriotti},\ and\ \citenamefont {Roke}}]{chen2016electrolytes}%
  \BibitemOpen
  \bibfield  {author} {\bibinfo {author} {\bibfnamefont {Y.}~\bibnamefont
  {Chen}}, \bibinfo {author} {\bibfnamefont {H.~I.}\ \bibnamefont {Okur}},
  \bibinfo {author} {\bibfnamefont {N.}~\bibnamefont {Gomopoulos}}, \bibinfo
  {author} {\bibfnamefont {C.}~\bibnamefont {Macias-Romero}}, \bibinfo {author}
  {\bibfnamefont {P.~S.}\ \bibnamefont {Cremer}}, \bibinfo {author}
  {\bibfnamefont {P.~B.}\ \bibnamefont {Petersen}}, \bibinfo {author}
  {\bibfnamefont {G.}~\bibnamefont {Tocci}}, \bibinfo {author} {\bibfnamefont
  {D.~M.}\ \bibnamefont {Wilkins}}, \bibinfo {author} {\bibfnamefont
  {C.}~\bibnamefont {Liang}}, \bibinfo {author} {\bibfnamefont
  {M.}~\bibnamefont {Ceriotti}}, \ and\ \bibinfo {author} {\bibfnamefont
  {S.}~\bibnamefont {Roke}},\ }\href@noop {} {\bibfield  {journal} {\bibinfo
  {journal} {Sci. Adv.}\ }\textbf {\bibinfo {volume} {2}},\ \bibinfo {pages}
  {e1501891} (\bibinfo {year} {2016})}\BibitemShut {NoStop}%
\bibitem [{\citenamefont {Hummer}, \citenamefont {Pratt},\ and\ \citenamefont
  {Garc\'{i}a}(1998)}]{HummerGarcia1998sjc}%
  \BibitemOpen
  \bibfield  {author} {\bibinfo {author} {\bibfnamefont {G.}~\bibnamefont
  {Hummer}}, \bibinfo {author} {\bibfnamefont {L.~R.}\ \bibnamefont {Pratt}}, \
  and\ \bibinfo {author} {\bibfnamefont {A.~E.}\ \bibnamefont {Garc\'{i}a}},\
  }\href {\doibase 10.1021/jp982195r} {\bibfield  {journal} {\bibinfo
  {journal} {J. Phys. Chem. A}\ }\textbf {\bibinfo {volume} {102}},\ \bibinfo
  {pages} {7885} (\bibinfo {year} {1998})}\BibitemShut {NoStop}%
\bibitem [{\citenamefont {Lum}, \citenamefont {Chandler},\ and\ \citenamefont
  {Weeks}(1999)}]{LumWeeks1999sjc}%
  \BibitemOpen
  \bibfield  {author} {\bibinfo {author} {\bibfnamefont {K.}~\bibnamefont
  {Lum}}, \bibinfo {author} {\bibfnamefont {D.}~\bibnamefont {Chandler}}, \
  and\ \bibinfo {author} {\bibfnamefont {J.~D.}\ \bibnamefont {Weeks}},\ }\href
  {\doibase 10.1021/jp984327m} {\bibfield  {journal} {\bibinfo  {journal} {J.
  Phys. Chem. B}\ }\textbf {\bibinfo {volume} {103}},\ \bibinfo {pages} {4570}
  (\bibinfo {year} {1999})}\BibitemShut {NoStop}%
\bibitem [{\citenamefont {Ashbaugh}(2000)}]{ashbaugh2000convergence}%
  \BibitemOpen
  \bibfield  {author} {\bibinfo {author} {\bibfnamefont {H.~S.}\ \bibnamefont
  {Ashbaugh}},\ }\href@noop {} {\bibfield  {journal} {\bibinfo  {journal} {J.
  Phys. Chem. B}\ }\textbf {\bibinfo {volume} {104}},\ \bibinfo {pages} {7235}
  (\bibinfo {year} {2000})}\BibitemShut {NoStop}%
\bibitem [{\citenamefont {Horinek}\ and\ \citenamefont
  {Netz}(2007)}]{horinek2007specific}%
  \BibitemOpen
  \bibfield  {author} {\bibinfo {author} {\bibfnamefont {D.}~\bibnamefont
  {Horinek}}\ and\ \bibinfo {author} {\bibfnamefont {R.~R.}\ \bibnamefont
  {Netz}},\ }\href@noop {} {\bibfield  {journal} {\bibinfo  {journal} {Phys.
  Rev. Lett.}\ }\textbf {\bibinfo {volume} {99}},\ \bibinfo {pages} {226104}
  (\bibinfo {year} {2007})}\BibitemShut {NoStop}%
\bibitem [{\citenamefont {Levin}(2009)}]{levin2009polarizable}%
  \BibitemOpen
  \bibfield  {author} {\bibinfo {author} {\bibfnamefont {Y.}~\bibnamefont
  {Levin}},\ }\href@noop {} {\bibfield  {journal} {\bibinfo  {journal} {Phys.
  Rev. Lett.}\ }\textbf {\bibinfo {volume} {102}},\ \bibinfo {pages} {147803}
  (\bibinfo {year} {2009})}\BibitemShut {NoStop}%
\bibitem [{\citenamefont {Ben-Amotz}(2016)}]{ben2016interfacial}%
  \BibitemOpen
  \bibfield  {author} {\bibinfo {author} {\bibfnamefont {D.}~\bibnamefont
  {Ben-Amotz}},\ }\href@noop {} {\bibfield  {journal} {\bibinfo  {journal} {J.
  Phys.: Condens. Matter}\ }\textbf {\bibinfo {volume} {28}},\ \bibinfo {pages}
  {414013} (\bibinfo {year} {2016})}\BibitemShut {NoStop}%
\bibitem [{\citenamefont {Beck}(2013)}]{beck2013influence}%
  \BibitemOpen
  \bibfield  {author} {\bibinfo {author} {\bibfnamefont {T.~L.}\ \bibnamefont
  {Beck}},\ }\href@noop {} {\bibfield  {journal} {\bibinfo  {journal} {Chem.
  Phys. Lett.}\ }\textbf {\bibinfo {volume} {561}},\ \bibinfo {pages} {1}
  (\bibinfo {year} {2013})}\BibitemShut {NoStop}%
\bibitem [{\citenamefont {Remsing}\ \emph {et~al.}(2014)\citenamefont
  {Remsing}, \citenamefont {Baer}, \citenamefont {Schenter}, \citenamefont
  {Mundy},\ and\ \citenamefont {Weeks}}]{remsing2014role}%
  \BibitemOpen
  \bibfield  {author} {\bibinfo {author} {\bibfnamefont {R.~C.}\ \bibnamefont
  {Remsing}}, \bibinfo {author} {\bibfnamefont {M.~D.}\ \bibnamefont {Baer}},
  \bibinfo {author} {\bibfnamefont {G.~K.}\ \bibnamefont {Schenter}}, \bibinfo
  {author} {\bibfnamefont {C.~J.}\ \bibnamefont {Mundy}}, \ and\ \bibinfo
  {author} {\bibfnamefont {J.~D.}\ \bibnamefont {Weeks}},\ }\href@noop {}
  {\bibfield  {journal} {\bibinfo  {journal} {J. Phys. Chem. Lett.}\ }\textbf
  {\bibinfo {volume} {5}},\ \bibinfo {pages} {2767} (\bibinfo {year}
  {2014})}\BibitemShut {NoStop}%
\bibitem [{\citenamefont {Cox}\ \emph {et~al.}(2020)\citenamefont {Cox},
  \citenamefont {Thorpe}, \citenamefont {Shaffer},\ and\ \citenamefont
  {Geissler}}]{cox2020assessing}%
  \BibitemOpen
  \bibfield  {author} {\bibinfo {author} {\bibfnamefont {S.~J.}\ \bibnamefont
  {Cox}}, \bibinfo {author} {\bibfnamefont {D.~G.}\ \bibnamefont {Thorpe}},
  \bibinfo {author} {\bibfnamefont {P.~R.}\ \bibnamefont {Shaffer}}, \ and\
  \bibinfo {author} {\bibfnamefont {P.~L.}\ \bibnamefont {Geissler}},\
  }\href@noop {} {\bibfield  {journal} {\bibinfo  {journal} {Chem. Sci.}\
  }\textbf {\bibinfo {volume} {11}},\ \bibinfo {pages} {11791} (\bibinfo {year}
  {2020})}\BibitemShut {NoStop}%
\bibitem [{\citenamefont {Born}(1920)}]{born1920volumen}%
  \BibitemOpen
  \bibfield  {author} {\bibinfo {author} {\bibfnamefont {M.}~\bibnamefont
  {Born}},\ }\href@noop {} {\bibfield  {journal} {\bibinfo  {journal}
  {Zeitschrift f{\"u}r Physik}\ }\textbf {\bibinfo {volume} {1}},\ \bibinfo
  {pages} {45} (\bibinfo {year} {1920})}\BibitemShut {NoStop}%
\bibitem [{\citenamefont {Berendsen}, \citenamefont {Grigera},\ and\
  \citenamefont {Straatsma}(1987)}]{BerendsenStraatsma1987sjc}%
  \BibitemOpen
  \bibfield  {author} {\bibinfo {author} {\bibfnamefont {H.~J.~C.}\
  \bibnamefont {Berendsen}}, \bibinfo {author} {\bibfnamefont {J.~R.}\
  \bibnamefont {Grigera}}, \ and\ \bibinfo {author} {\bibfnamefont {T.~P.}\
  \bibnamefont {Straatsma}},\ }\href@noop {} {\bibfield  {journal} {\bibinfo
  {journal} {J. Phys. Chem.}\ }\textbf {\bibinfo {volume} {91}},\ \bibinfo
  {pages} {6269} (\bibinfo {year} {1987})}\BibitemShut {NoStop}%
\bibitem [{\citenamefont {Hansen}\ and\ \citenamefont
  {McDonald}(2013)}]{HANSEN2013265}%
  \BibitemOpen
  \bibfield  {author} {\bibinfo {author} {\bibfnamefont {J.-P.}\ \bibnamefont
  {Hansen}}\ and\ \bibinfo {author} {\bibfnamefont {I.~R.}\ \bibnamefont
  {McDonald}},\ }\href {\doibase
  https://doi.org/10.1016/B978-0-12-387032-2.00007-6} {\emph {\bibinfo {title}
  {Theory of Simple Liquids}}},\ \bibinfo {edition} {4th}\ ed.\ (\bibinfo
  {publisher} {Academic Press},\ \bibinfo {year} {2013})\BibitemShut {NoStop}%
\bibitem [{\citenamefont {Tomasi}, \citenamefont {Mennucci},\ and\
  \citenamefont {Cammi}(2005)}]{tomasi2005quantum}%
  \BibitemOpen
  \bibfield  {author} {\bibinfo {author} {\bibfnamefont {J.}~\bibnamefont
  {Tomasi}}, \bibinfo {author} {\bibfnamefont {B.}~\bibnamefont {Mennucci}}, \
  and\ \bibinfo {author} {\bibfnamefont {R.}~\bibnamefont {Cammi}},\
  }\href@noop {} {\bibfield  {journal} {\bibinfo  {journal} {Chem. Rev.}\
  }\textbf {\bibinfo {volume} {105}},\ \bibinfo {pages} {2999} (\bibinfo {year}
  {2005})}\BibitemShut {NoStop}%
\bibitem [{\citenamefont {Latimer}, \citenamefont {Pitzer},\ and\ \citenamefont
  {Slansky}(1939)}]{latimer1939free}%
  \BibitemOpen
  \bibfield  {author} {\bibinfo {author} {\bibfnamefont {W.~M.}\ \bibnamefont
  {Latimer}}, \bibinfo {author} {\bibfnamefont {K.~S.}\ \bibnamefont {Pitzer}},
  \ and\ \bibinfo {author} {\bibfnamefont {C.~M.}\ \bibnamefont {Slansky}},\
  }\href@noop {} {\bibfield  {journal} {\bibinfo  {journal} {J. Chem. Phys.}\
  }\textbf {\bibinfo {volume} {7}},\ \bibinfo {pages} {108} (\bibinfo {year}
  {1939})}\BibitemShut {NoStop}%
\bibitem [{\citenamefont {Rajamani}, \citenamefont {Ghosh},\ and\ \citenamefont
  {Garde}(2004)}]{rajamani2004size}%
  \BibitemOpen
  \bibfield  {author} {\bibinfo {author} {\bibfnamefont {S.}~\bibnamefont
  {Rajamani}}, \bibinfo {author} {\bibfnamefont {T.}~\bibnamefont {Ghosh}}, \
  and\ \bibinfo {author} {\bibfnamefont {S.}~\bibnamefont {Garde}},\
  }\href@noop {} {\bibfield  {journal} {\bibinfo  {journal} {J. Chem. Phys.}\
  }\textbf {\bibinfo {volume} {120}},\ \bibinfo {pages} {4457} (\bibinfo {year}
  {2004})}\BibitemShut {NoStop}%
\bibitem [{\citenamefont {Bardhan}, \citenamefont {Jungwirth},\ and\
  \citenamefont {Makowski}(2012)}]{bardhan2012affine}%
  \BibitemOpen
  \bibfield  {author} {\bibinfo {author} {\bibfnamefont {J.~P.}\ \bibnamefont
  {Bardhan}}, \bibinfo {author} {\bibfnamefont {P.}~\bibnamefont {Jungwirth}},
  \ and\ \bibinfo {author} {\bibfnamefont {L.}~\bibnamefont {Makowski}},\
  }\href@noop {} {\bibfield  {journal} {\bibinfo  {journal} {J. Chem. Phys.}\
  }\textbf {\bibinfo {volume} {137}},\ \bibinfo {pages} {124101} (\bibinfo
  {year} {2012})}\BibitemShut {NoStop}%
\bibitem [{\citenamefont {Marcus}(1956)}]{marcus1956theoryI}%
  \BibitemOpen
  \bibfield  {author} {\bibinfo {author} {\bibfnamefont {R.~A.}\ \bibnamefont
  {Marcus}},\ }\href@noop {} {\bibfield  {journal} {\bibinfo  {journal} {J.
  Chem. Phys.}\ }\textbf {\bibinfo {volume} {24}},\ \bibinfo {pages} {966}
  (\bibinfo {year} {1956})}\BibitemShut {NoStop}%
\bibitem [{\citenamefont {Hummer}, \citenamefont {Pratt},\ and\ \citenamefont
  {Garc\'{i}a}(1997)}]{HummerGarcia1997sjc}%
  \BibitemOpen
  \bibfield  {author} {\bibinfo {author} {\bibfnamefont {G.}~\bibnamefont
  {Hummer}}, \bibinfo {author} {\bibfnamefont {L.~R.}\ \bibnamefont {Pratt}}, \
  and\ \bibinfo {author} {\bibfnamefont {A.~E.}\ \bibnamefont {Garc\'{i}a}},\
  }\href {\doibase 10.1021/ja971148u} {\bibfield  {journal} {\bibinfo
  {journal} {J. Am. Chem. Soc.}\ }\textbf {\bibinfo {volume} {119}},\ \bibinfo
  {pages} {8523} (\bibinfo {year} {1997})}\BibitemShut {NoStop}%
\bibitem [{\citenamefont {Pratt}\ and\ \citenamefont
  {LaViolette}(1998)}]{pratt1998quasi}%
  \BibitemOpen
  \bibfield  {author} {\bibinfo {author} {\bibfnamefont {L.~R.}\ \bibnamefont
  {Pratt}}\ and\ \bibinfo {author} {\bibfnamefont {R.~A.}\ \bibnamefont
  {LaViolette}},\ }\href@noop {} {\bibfield  {journal} {\bibinfo  {journal}
  {Mol. Phys.}\ }\textbf {\bibinfo {volume} {94}},\ \bibinfo {pages} {909}
  (\bibinfo {year} {1998})}\BibitemShut {NoStop}%
\bibitem [{\citenamefont {Beck}, \citenamefont {Paulaitis},\ and\ \citenamefont
  {Pratt}(2012)}]{2012TPDT}%
  \BibitemOpen
  \bibfield  {author} {\bibinfo {author} {\bibfnamefont {T.~L.}\ \bibnamefont
  {Beck}}, \bibinfo {author} {\bibfnamefont {M.~E.}\ \bibnamefont {Paulaitis}},
  \ and\ \bibinfo {author} {\bibfnamefont {L.~R.}\ \bibnamefont {Pratt}},\
  }\href@noop {} {\emph {\bibinfo {title} {The Potential Distribution Theorem
  and Models of Molecular Solutions}}}\ (\bibinfo  {publisher} {Cambridge
  University Press},\ \bibinfo {address} {Cambridge, United Kingdom},\ \bibinfo
  {year} {2012})\BibitemShut {NoStop}%
\bibitem [{\citenamefont {Duignan}\ \emph
  {et~al.}(2017{\natexlab{a}})\citenamefont {Duignan}, \citenamefont {Baer},
  \citenamefont {Schenter},\ and\ \citenamefont {Mundy}}]{duignan2017real}%
  \BibitemOpen
  \bibfield  {author} {\bibinfo {author} {\bibfnamefont {T.~T.}\ \bibnamefont
  {Duignan}}, \bibinfo {author} {\bibfnamefont {M.~D.}\ \bibnamefont {Baer}},
  \bibinfo {author} {\bibfnamefont {G.~K.}\ \bibnamefont {Schenter}}, \ and\
  \bibinfo {author} {\bibfnamefont {C.~J.}\ \bibnamefont {Mundy}},\ }\href@noop
  {} {\bibfield  {journal} {\bibinfo  {journal} {Chem. Sci.}\ }\textbf
  {\bibinfo {volume} {8}},\ \bibinfo {pages} {6131} (\bibinfo {year}
  {2017}{\natexlab{a}})}\BibitemShut {NoStop}%
\bibitem [{\citenamefont {Duignan}\ \emph
  {et~al.}(2017{\natexlab{b}})\citenamefont {Duignan}, \citenamefont {Baer},
  \citenamefont {Schenter},\ and\ \citenamefont
  {Mundy}}]{duignan2017electrostatic}%
  \BibitemOpen
  \bibfield  {author} {\bibinfo {author} {\bibfnamefont {T.~T.}\ \bibnamefont
  {Duignan}}, \bibinfo {author} {\bibfnamefont {M.~D.}\ \bibnamefont {Baer}},
  \bibinfo {author} {\bibfnamefont {G.~K.}\ \bibnamefont {Schenter}}, \ and\
  \bibinfo {author} {\bibfnamefont {C.~J.}\ \bibnamefont {Mundy}},\ }\href
  {\doibase 10.1063/1.4994912} {\bibfield  {journal} {\bibinfo  {journal} {J.
  Chem. Phys.}\ }\textbf {\bibinfo {volume} {147}},\ \bibinfo {pages} {161716}
  (\bibinfo {year} {2017}{\natexlab{b}})}\BibitemShut {NoStop}%
\bibitem [{\citenamefont {Song}, \citenamefont {Chandler},\ and\ \citenamefont
  {Marcus}(1996)}]{song1996gaussian}%
  \BibitemOpen
  \bibfield  {author} {\bibinfo {author} {\bibfnamefont {X.}~\bibnamefont
  {Song}}, \bibinfo {author} {\bibfnamefont {D.}~\bibnamefont {Chandler}}, \
  and\ \bibinfo {author} {\bibfnamefont {R.}~\bibnamefont {Marcus}},\
  }\href@noop {} {\bibfield  {journal} {\bibinfo  {journal} {J. Phys. Chem.}\
  }\textbf {\bibinfo {volume} {100}},\ \bibinfo {pages} {11954} (\bibinfo
  {year} {1996})}\BibitemShut {NoStop}%
\bibitem [{\citenamefont {Madden}\ and\ \citenamefont
  {Kivelson}(1984)}]{MaddenKivelson1984sjc}%
  \BibitemOpen
  \bibfield  {author} {\bibinfo {author} {\bibfnamefont {P.}~\bibnamefont
  {Madden}}\ and\ \bibinfo {author} {\bibfnamefont {D.}~\bibnamefont
  {Kivelson}},\ }\enquote {\bibinfo {title} {A consistent molecular treatment
  of dielectric phenomena},}\ in\ \href {\doibase 10.1002/9780470142806.ch5}
  {\emph {\bibinfo {booktitle} {Adv. Chem. Phys.}}}\ (\bibinfo  {publisher}
  {John Wiley \& Sons, Inc.},\ \bibinfo {year} {1984})\ pp.\ \bibinfo {pages}
  {467--566}\BibitemShut {NoStop}%
\bibitem [{\citenamefont {Ballenegger}\ and\ \citenamefont
  {Hansen}(2005)}]{ballenegger2005dielectric}%
  \BibitemOpen
  \bibfield  {author} {\bibinfo {author} {\bibfnamefont {V.}~\bibnamefont
  {Ballenegger}}\ and\ \bibinfo {author} {\bibfnamefont {J.-P.}\ \bibnamefont
  {Hansen}},\ }\href@noop {} {\bibfield  {journal} {\bibinfo  {journal} {J.
  Chem. Phys}\ }\textbf {\bibinfo {volume} {122}},\ \bibinfo {pages} {114711}
  (\bibinfo {year} {2005})}\BibitemShut {NoStop}%
\bibitem [{\citenamefont {Schlaich}, \citenamefont {Knapp},\ and\ \citenamefont
  {Netz}(2016)}]{schlaich2016water}%
  \BibitemOpen
  \bibfield  {author} {\bibinfo {author} {\bibfnamefont {A.}~\bibnamefont
  {Schlaich}}, \bibinfo {author} {\bibfnamefont {E.~W.}\ \bibnamefont {Knapp}},
  \ and\ \bibinfo {author} {\bibfnamefont {R.~R.}\ \bibnamefont {Netz}},\
  }\href@noop {} {\bibfield  {journal} {\bibinfo  {journal} {Phys. Rev. Lett.}\
  }\textbf {\bibinfo {volume} {117}},\ \bibinfo {pages} {048001} (\bibinfo
  {year} {2016})}\BibitemShut {NoStop}%
\bibitem [{\citenamefont {Loche}\ \emph {et~al.}(2018)\citenamefont {Loche},
  \citenamefont {Ayaz}, \citenamefont {Schlaich}, \citenamefont {Bonthuis},\
  and\ \citenamefont {Netz}}]{loche2018breakdown}%
  \BibitemOpen
  \bibfield  {author} {\bibinfo {author} {\bibfnamefont {P.}~\bibnamefont
  {Loche}}, \bibinfo {author} {\bibfnamefont {C.}~\bibnamefont {Ayaz}},
  \bibinfo {author} {\bibfnamefont {A.}~\bibnamefont {Schlaich}}, \bibinfo
  {author} {\bibfnamefont {D.~J.}\ \bibnamefont {Bonthuis}}, \ and\ \bibinfo
  {author} {\bibfnamefont {R.~R.}\ \bibnamefont {Netz}},\ }\href@noop {}
  {\bibfield  {journal} {\bibinfo  {journal} {J. Phys. Chem. Lett.}\ }\textbf
  {\bibinfo {volume} {9}},\ \bibinfo {pages} {6463} (\bibinfo {year}
  {2018})}\BibitemShut {NoStop}%
\bibitem [{\citenamefont {Zhang}\ and\ \citenamefont
  {Sprik}(2020)}]{zhang2020electromechanics}%
  \BibitemOpen
  \bibfield  {author} {\bibinfo {author} {\bibfnamefont {C.}~\bibnamefont
  {Zhang}}\ and\ \bibinfo {author} {\bibfnamefont {M.}~\bibnamefont {Sprik}},\
  }\href@noop {} {\bibfield  {journal} {\bibinfo  {journal} {Phys. Chem. Chem.
  Phys.}\ }\textbf {\bibinfo {volume} {22}},\ \bibinfo {pages} {10676}
  (\bibinfo {year} {2020})}\BibitemShut {NoStop}%
\bibitem [{\citenamefont {Hummer}, \citenamefont {Pratt},\ and\ \citenamefont
  {Garc\'{i}a}(1996)}]{HummerGarcia1996sjc}%
  \BibitemOpen
  \bibfield  {author} {\bibinfo {author} {\bibfnamefont {G.}~\bibnamefont
  {Hummer}}, \bibinfo {author} {\bibfnamefont {L.~R.}\ \bibnamefont {Pratt}}, \
  and\ \bibinfo {author} {\bibfnamefont {A.~E.}\ \bibnamefont {Garc\'{i}a}},\
  }\href {\doibase 10.1021/jp951011v} {\bibfield  {journal} {\bibinfo
  {journal} {J. Phys. Chem.}\ }\textbf {\bibinfo {volume} {100}},\ \bibinfo
  {pages} {1206} (\bibinfo {year} {1996})}\BibitemShut {NoStop}%
\bibitem [{\citenamefont {{\AA}qvist}\ and\ \citenamefont
  {Hansson}(1998)}]{aaqvist1998analysis}%
  \BibitemOpen
  \bibfield  {author} {\bibinfo {author} {\bibfnamefont {J.}~\bibnamefont
  {{\AA}qvist}}\ and\ \bibinfo {author} {\bibfnamefont {T.}~\bibnamefont
  {Hansson}},\ }\href@noop {} {\bibfield  {journal} {\bibinfo  {journal} {J.
  Phys. Chem. B}\ }\textbf {\bibinfo {volume} {102}},\ \bibinfo {pages} {3837}
  (\bibinfo {year} {1998})}\BibitemShut {NoStop}%
\bibitem [{\citenamefont {Harder}\ and\ \citenamefont
  {Roux}(2008)}]{harder2008origin}%
  \BibitemOpen
  \bibfield  {author} {\bibinfo {author} {\bibfnamefont {E.}~\bibnamefont
  {Harder}}\ and\ \bibinfo {author} {\bibfnamefont {B.}~\bibnamefont {Roux}},\
  }\href@noop {} {\bibfield  {journal} {\bibinfo  {journal} {J. Chem. Phys.}\
  }\textbf {\bibinfo {volume} {129}},\ \bibinfo {pages} {234706} (\bibinfo
  {year} {2008})}\BibitemShut {NoStop}%
\bibitem [{\citenamefont {Arslanargin}\ and\ \citenamefont
  {Beck}(2012)}]{arslanargin2012free}%
  \BibitemOpen
  \bibfield  {author} {\bibinfo {author} {\bibfnamefont {A.}~\bibnamefont
  {Arslanargin}}\ and\ \bibinfo {author} {\bibfnamefont {T.~L.}\ \bibnamefont
  {Beck}},\ }\href@noop {} {\bibfield  {journal} {\bibinfo  {journal} {J. Chem.
  Phys.}\ }\textbf {\bibinfo {volume} {136}},\ \bibinfo {pages} {104503}
  (\bibinfo {year} {2012})}\BibitemShut {NoStop}%
\bibitem [{\citenamefont {Horv{\'a}th}\ \emph {et~al.}(2013)\citenamefont
  {Horv{\'a}th}, \citenamefont {Beu}, \citenamefont {Manghi},\ and\
  \citenamefont {Palmeri}}]{horvath2013vapor}%
  \BibitemOpen
  \bibfield  {author} {\bibinfo {author} {\bibfnamefont {L.}~\bibnamefont
  {Horv{\'a}th}}, \bibinfo {author} {\bibfnamefont {T.}~\bibnamefont {Beu}},
  \bibinfo {author} {\bibfnamefont {M.}~\bibnamefont {Manghi}}, \ and\ \bibinfo
  {author} {\bibfnamefont {J.}~\bibnamefont {Palmeri}},\ }\href@noop {}
  {\bibfield  {journal} {\bibinfo  {journal} {J. Chem. Phys.}\ }\textbf
  {\bibinfo {volume} {138}},\ \bibinfo {pages} {154702} (\bibinfo {year}
  {2013})}\BibitemShut {NoStop}%
\bibitem [{\citenamefont {Remsing}\ and\ \citenamefont
  {Weeks}(2016)}]{remsing2016role}%
  \BibitemOpen
  \bibfield  {author} {\bibinfo {author} {\bibfnamefont {R.~C.}\ \bibnamefont
  {Remsing}}\ and\ \bibinfo {author} {\bibfnamefont {J.~D.}\ \bibnamefont
  {Weeks}},\ }\href@noop {} {\bibfield  {journal} {\bibinfo  {journal} {J.
  Phys. Chem. B}\ }\textbf {\bibinfo {volume} {120}},\ \bibinfo {pages} {6238}
  (\bibinfo {year} {2016})}\BibitemShut {NoStop}%
\bibitem [{\citenamefont {Doyle}, \citenamefont {Shi},\ and\ \citenamefont
  {Beck}(2019)}]{doyle2019importance}%
  \BibitemOpen
  \bibfield  {author} {\bibinfo {author} {\bibfnamefont {C.~C.}\ \bibnamefont
  {Doyle}}, \bibinfo {author} {\bibfnamefont {Y.}~\bibnamefont {Shi}}, \ and\
  \bibinfo {author} {\bibfnamefont {T.~L.}\ \bibnamefont {Beck}},\ }\href@noop
  {} {\bibfield  {journal} {\bibinfo  {journal} {J. Phys. Chem. B}\ }\textbf
  {\bibinfo {volume} {123}},\ \bibinfo {pages} {3348} (\bibinfo {year}
  {2019})}\BibitemShut {NoStop}%
\bibitem [{\citenamefont {Lynden-Bell}\ and\ \citenamefont
  {Rasaiah}(1997)}]{lynden1997hydrophobic}%
  \BibitemOpen
  \bibfield  {author} {\bibinfo {author} {\bibfnamefont {R.}~\bibnamefont
  {Lynden-Bell}}\ and\ \bibinfo {author} {\bibfnamefont {J.}~\bibnamefont
  {Rasaiah}},\ }\href@noop {} {\bibfield  {journal} {\bibinfo  {journal} {J.
  Chem. Phys.}\ }\textbf {\bibinfo {volume} {107}},\ \bibinfo {pages} {1981}
  (\bibinfo {year} {1997})}\BibitemShut {NoStop}%
\bibitem [{\citenamefont {Figueirido}, \citenamefont {Del~Buono},\ and\
  \citenamefont {Levy}(1995)}]{FigueiridoLevy1995sjc}%
  \BibitemOpen
  \bibfield  {author} {\bibinfo {author} {\bibfnamefont {F.}~\bibnamefont
  {Figueirido}}, \bibinfo {author} {\bibfnamefont {G.~S.}\ \bibnamefont
  {Del~Buono}}, \ and\ \bibinfo {author} {\bibfnamefont {R.~M.}\ \bibnamefont
  {Levy}},\ }\href {\doibase 10.1063/1.470721} {\bibfield  {journal} {\bibinfo
  {journal} {J. Chem. Phys.}\ }\textbf {\bibinfo {volume} {103}},\ \bibinfo
  {pages} {6133} (\bibinfo {year} {1995})}\BibitemShut {NoStop}%
\bibitem [{\citenamefont {H\"{u}nenberger}\ and\ \citenamefont
  {McCammon}(1999)}]{HunenbergerMcCammon1999sjc}%
  \BibitemOpen
  \bibfield  {author} {\bibinfo {author} {\bibfnamefont {P.~H.}\ \bibnamefont
  {H\"{u}nenberger}}\ and\ \bibinfo {author} {\bibfnamefont {J.~A.}\
  \bibnamefont {McCammon}},\ }\href {\doibase 10.1063/1.477873} {\bibfield
  {journal} {\bibinfo  {journal} {J. Chem. Phys.}\ }\textbf {\bibinfo {volume}
  {110}},\ \bibinfo {pages} {1856} (\bibinfo {year} {1999})}\BibitemShut
  {NoStop}%
\bibitem [{\citenamefont {Cox}\ and\ \citenamefont
  {Geissler}(2018)}]{cox2018interfacial}%
  \BibitemOpen
  \bibfield  {author} {\bibinfo {author} {\bibfnamefont {S.~J.}\ \bibnamefont
  {Cox}}\ and\ \bibinfo {author} {\bibfnamefont {P.~L.}\ \bibnamefont
  {Geissler}},\ }\href {\doibase 10.1063/1.5020563} {\bibfield  {journal}
  {\bibinfo  {journal} {J. Chem. Phys.}\ }\textbf {\bibinfo {volume} {148}},\
  \bibinfo {pages} {222823} (\bibinfo {year} {2018})}\BibitemShut {NoStop}%
\bibitem [{\citenamefont {Cox}(2020)}]{cox2020dielectric}%
  \BibitemOpen
  \bibfield  {author} {\bibinfo {author} {\bibfnamefont {S.~J.}\ \bibnamefont
  {Cox}},\ }\href@noop {} {\bibfield  {journal} {\bibinfo  {journal} {Proc.
  Natl. Acad. Sci.}\ }\textbf {\bibinfo {volume} {117}},\ \bibinfo {pages}
  {19746} (\bibinfo {year} {2020})}\BibitemShut {NoStop}%
\bibitem [{\citenamefont {Stillinger}\ and\ \citenamefont
  {Rahman}(1974)}]{stillinger1974improved}%
  \BibitemOpen
  \bibfield  {author} {\bibinfo {author} {\bibfnamefont {F.~H.}\ \bibnamefont
  {Stillinger}}\ and\ \bibinfo {author} {\bibfnamefont {A.}~\bibnamefont
  {Rahman}},\ }\href@noop {} {\bibfield  {journal} {\bibinfo  {journal} {J.
  Chem. Phys.}\ }\textbf {\bibinfo {volume} {60}},\ \bibinfo {pages} {1545}
  (\bibinfo {year} {1974})}\BibitemShut {NoStop}%
\bibitem [{\citenamefont {Chandler}(1993)}]{Chandler1993sjc}%
  \BibitemOpen
  \bibfield  {author} {\bibinfo {author} {\bibfnamefont {D.}~\bibnamefont
  {Chandler}},\ }\href@noop {} {\bibfield  {journal} {\bibinfo  {journal}
  {Phys. Rev. E}\ }\textbf {\bibinfo {volume} {48}},\ \bibinfo {pages} {2898}
  (\bibinfo {year} {1993})}\BibitemShut {NoStop}%
\bibitem [{Note1()}]{Note1}%
  \BibitemOpen
  \bibinfo {note} {We again equate $R_0$ with the Lennard-Jones diameter for
  ion-water interactions. An alternative estimate of $R_0$ from the first peak
  of ion-oxygen radial distribution functions yields a similar relationship
  between $R$ and $R_0$, but with a smaller offset.}\BibitemShut {Stop}%
\bibitem [{\citenamefont {Mukhopadhyay}\ \emph {et~al.}(2012)\citenamefont
  {Mukhopadhyay}, \citenamefont {Fenley}, \citenamefont {Tolokh},\ and\
  \citenamefont {Onufriev}}]{MukhopadhyayOnufriev2012sjc}%
  \BibitemOpen
  \bibfield  {author} {\bibinfo {author} {\bibfnamefont {A.}~\bibnamefont
  {Mukhopadhyay}}, \bibinfo {author} {\bibfnamefont {A.~T.}\ \bibnamefont
  {Fenley}}, \bibinfo {author} {\bibfnamefont {I.~S.}\ \bibnamefont {Tolokh}},
  \ and\ \bibinfo {author} {\bibfnamefont {A.~V.}\ \bibnamefont {Onufriev}},\
  }\href {\doibase 10.1021/jp305226j} {\bibfield  {journal} {\bibinfo
  {journal} {J. Phys. Chem. B}\ }\textbf {\bibinfo {volume} {116}},\ \bibinfo
  {pages} {9776} (\bibinfo {year} {2012})}\BibitemShut {NoStop}%
\bibitem [{\citenamefont {Mukhopadhyay}\ \emph {et~al.}(2014)\citenamefont
  {Mukhopadhyay}, \citenamefont {Aguilar}, \citenamefont {Tolokh},\ and\
  \citenamefont {Onufriev}}]{mukhopadhyay2014introducing}%
  \BibitemOpen
  \bibfield  {author} {\bibinfo {author} {\bibfnamefont {A.}~\bibnamefont
  {Mukhopadhyay}}, \bibinfo {author} {\bibfnamefont {B.~H.}\ \bibnamefont
  {Aguilar}}, \bibinfo {author} {\bibfnamefont {I.~S.}\ \bibnamefont {Tolokh}},
  \ and\ \bibinfo {author} {\bibfnamefont {A.~V.}\ \bibnamefont {Onufriev}},\
  }\href@noop {} {\bibfield  {journal} {\bibinfo  {journal} {J. Chem. Theor.
  Comput.}\ }\textbf {\bibinfo {volume} {10}},\ \bibinfo {pages} {1788}
  (\bibinfo {year} {2014})}\BibitemShut {NoStop}%
\bibitem [{\citenamefont {Dinpajooh}\ and\ \citenamefont
  {Matyushov}(2015)}]{dinpajooh2015free}%
  \BibitemOpen
  \bibfield  {author} {\bibinfo {author} {\bibfnamefont {M.}~\bibnamefont
  {Dinpajooh}}\ and\ \bibinfo {author} {\bibfnamefont {D.~V.}\ \bibnamefont
  {Matyushov}},\ }\href@noop {} {\bibfield  {journal} {\bibinfo  {journal} {J.
  Chem. Phys.}\ }\textbf {\bibinfo {volume} {143}},\ \bibinfo {pages} {044511}
  (\bibinfo {year} {2015})}\BibitemShut {NoStop}%
\bibitem [{\citenamefont {Duignan}\ and\ \citenamefont
  {Zhao}(2020)}]{duignan2020born}%
  \BibitemOpen
  \bibfield  {author} {\bibinfo {author} {\bibfnamefont {T.~T.}\ \bibnamefont
  {Duignan}}\ and\ \bibinfo {author} {\bibfnamefont {X.~S.}\ \bibnamefont
  {Zhao}},\ }\href@noop {} {\bibfield  {journal} {\bibinfo  {journal} {Phys.
  Chem. Chem. Phys.}\ }\textbf {\bibinfo {volume} {22}},\ \bibinfo {pages}
  {25126} (\bibinfo {year} {2020})}\BibitemShut {NoStop}%
\bibitem [{\citenamefont {Pratt}, \citenamefont {Chaudhari},\ and\
  \citenamefont {Rempe}(2016)}]{pratt2016statistical}%
  \BibitemOpen
  \bibfield  {author} {\bibinfo {author} {\bibfnamefont {L.~R.}\ \bibnamefont
  {Pratt}}, \bibinfo {author} {\bibfnamefont {M.~I.}\ \bibnamefont
  {Chaudhari}}, \ and\ \bibinfo {author} {\bibfnamefont {S.~B.}\ \bibnamefont
  {Rempe}},\ }\href@noop {} {\bibfield  {journal} {\bibinfo  {journal} {J.
  Phys. Chem. B}\ }\textbf {\bibinfo {volume} {120}},\ \bibinfo {pages} {6455}
  (\bibinfo {year} {2016})}\BibitemShut {NoStop}%
\bibitem [{\citenamefont {Chandler}(2005)}]{chandler2005interfaces}%
  \BibitemOpen
  \bibfield  {author} {\bibinfo {author} {\bibfnamefont {D.}~\bibnamefont
  {Chandler}},\ }\href@noop {} {\bibfield  {journal} {\bibinfo  {journal}
  {Nature}\ }\textbf {\bibinfo {volume} {437}},\ \bibinfo {pages} {640}
  (\bibinfo {year} {2005})}\BibitemShut {NoStop}%
\bibitem [{\citenamefont {Varilly}, \citenamefont {Patel},\ and\ \citenamefont
  {Chandler}(2011)}]{VarillyChandler2011sjc}%
  \BibitemOpen
  \bibfield  {author} {\bibinfo {author} {\bibfnamefont {P.}~\bibnamefont
  {Varilly}}, \bibinfo {author} {\bibfnamefont {A.~J.}\ \bibnamefont {Patel}},
  \ and\ \bibinfo {author} {\bibfnamefont {D.}~\bibnamefont {Chandler}},\
  }\href@noop {} {\bibfield  {journal} {\bibinfo  {journal} {J. Chem. Phys.}\
  }\textbf {\bibinfo {volume} {134}},\ \bibinfo {pages} {074109} (\bibinfo
  {year} {2011})}\BibitemShut {NoStop}%
\bibitem [{\citenamefont {Vaikuntanathan}\ and\ \citenamefont
  {Geissler}(2014)}]{VaikuntanathanGeissler2014sjc}%
  \BibitemOpen
  \bibfield  {author} {\bibinfo {author} {\bibfnamefont {S.}~\bibnamefont
  {Vaikuntanathan}}\ and\ \bibinfo {author} {\bibfnamefont {P.~L.}\
  \bibnamefont {Geissler}},\ }\href@noop {} {\bibfield  {journal} {\bibinfo
  {journal} {Phys. Rev. Lett.}\ }\textbf {\bibinfo {volume} {112}},\ \bibinfo
  {pages} {020603} (\bibinfo {year} {2014})}\BibitemShut {NoStop}%
\bibitem [{\citenamefont {Vaikuntanathan}\ \emph {et~al.}(2016)\citenamefont
  {Vaikuntanathan}, \citenamefont {Rotskoff}, \citenamefont {Hudson},\ and\
  \citenamefont {Geissler}}]{VaikuntanathanGeissler2016sjc}%
  \BibitemOpen
  \bibfield  {author} {\bibinfo {author} {\bibfnamefont {S.}~\bibnamefont
  {Vaikuntanathan}}, \bibinfo {author} {\bibfnamefont {G.}~\bibnamefont
  {Rotskoff}}, \bibinfo {author} {\bibfnamefont {A.}~\bibnamefont {Hudson}}, \
  and\ \bibinfo {author} {\bibfnamefont {P.~L.}\ \bibnamefont {Geissler}},\
  }\href@noop {} {\bibfield  {journal} {\bibinfo  {journal} {Proc. Natl. Acad.
  Sci. USA}\ }\textbf {\bibinfo {volume} {113}},\ \bibinfo {pages} {E2224}
  (\bibinfo {year} {2016})}\BibitemShut {NoStop}%
\bibitem [{\citenamefont {Plimpton}(1995)}]{plimpton1995sjc}%
  \BibitemOpen
  \bibfield  {author} {\bibinfo {author} {\bibfnamefont {S.}~\bibnamefont
  {Plimpton}},\ }\href@noop {} {\bibfield  {journal} {\bibinfo  {journal} {J.
  Comput. Phys.}\ }\textbf {\bibinfo {volume} {117}},\ \bibinfo {pages} {1}
  (\bibinfo {year} {1995})}\BibitemShut {NoStop}%
\bibitem [{\citenamefont {Hockney}\ and\ \citenamefont
  {Eastwood}(1988)}]{HockneyEastwood1988sjc}%
  \BibitemOpen
  \bibfield  {author} {\bibinfo {author} {\bibfnamefont {R.~W.}\ \bibnamefont
  {Hockney}}\ and\ \bibinfo {author} {\bibfnamefont {J.~W.}\ \bibnamefont
  {Eastwood}},\ }\href@noop {} {\emph {\bibinfo {title} {Computer simulation
  using particles}}}\ (\bibinfo  {publisher} {CRC Press},\ \bibinfo {year}
  {1988})\BibitemShut {NoStop}%
\bibitem [{\citenamefont {Kolafa}\ and\ \citenamefont
  {Perram}(1992)}]{kolafa1992cutoff}%
  \BibitemOpen
  \bibfield  {author} {\bibinfo {author} {\bibfnamefont {J.}~\bibnamefont
  {Kolafa}}\ and\ \bibinfo {author} {\bibfnamefont {J.~W.}\ \bibnamefont
  {Perram}},\ }\href@noop {} {\bibfield  {journal} {\bibinfo  {journal} {Mol.
  Sim.}\ }\textbf {\bibinfo {volume} {9}},\ \bibinfo {pages} {351} (\bibinfo
  {year} {1992})}\BibitemShut {NoStop}%
\bibitem [{\citenamefont {Shirts}\ and\ \citenamefont
  {Chodera}(2008)}]{ShirtsChodera2008sjc}%
  \BibitemOpen
  \bibfield  {author} {\bibinfo {author} {\bibfnamefont {M.~R.}\ \bibnamefont
  {Shirts}}\ and\ \bibinfo {author} {\bibfnamefont {J.~D.}\ \bibnamefont
  {Chodera}},\ }\href {\doibase 10.1063/1.2978177} {\bibfield  {journal}
  {\bibinfo  {journal} {J. Chem. Phys.}\ }\textbf {\bibinfo {volume} {129}},\
  \bibinfo {pages} {124105} (\bibinfo {year} {2008})}\BibitemShut {NoStop}%
\bibitem [{\citenamefont {Branch}, \citenamefont {Coleman},\ and\ \citenamefont
  {Li}(1999)}]{branch1999subspace}%
  \BibitemOpen
  \bibfield  {author} {\bibinfo {author} {\bibfnamefont {M.~A.}\ \bibnamefont
  {Branch}}, \bibinfo {author} {\bibfnamefont {T.~F.}\ \bibnamefont {Coleman}},
  \ and\ \bibinfo {author} {\bibfnamefont {Y.}~\bibnamefont {Li}},\ }\href@noop
  {} {\bibfield  {journal} {\bibinfo  {journal} {{SIAM} J. Sci. Comput.}\
  }\textbf {\bibinfo {volume} {21}},\ \bibinfo {pages} {1} (\bibinfo {year}
  {1999})}\BibitemShut {NoStop}%
\bibitem [{\citenamefont {Vijaykumar}\ \emph {et~al.}(2020)\citenamefont
  {Vijaykumar}, \citenamefont {Bardelli}, \citenamefont {Rothberg},
  \citenamefont {Hilboll}, \citenamefont {Kloeckner}, \citenamefont {Scopatz},
  \citenamefont {Lee}, \citenamefont {Rokem}, \citenamefont {Woods},
  \citenamefont {Fulton}, \citenamefont {Masson}, \citenamefont
  {H\"{a}ggstr\"{o}m}, \citenamefont {Fitzgerald}, \citenamefont {Nicholson},
  \citenamefont {Hagen}, \citenamefont {Pasechnik}, \citenamefont {Olivetti},
  \citenamefont {Martin}, \citenamefont {Wieser}, \citenamefont {Silva},
  \citenamefont {Lenders}, \citenamefont {Wilhelm}, \citenamefont {Young},
  \citenamefont {Price}, \citenamefont {Ingold}, \citenamefont {Allen},
  \citenamefont {Lee}, \citenamefont {Audren}, \citenamefont {Probst},
  \citenamefont {Dietrich}, \citenamefont {Silterra}, \citenamefont {Webber},
  \citenamefont {Slavi\v{c}}, \citenamefont {Nothman}, \citenamefont {Buchner},
  \citenamefont {Kulick}, \citenamefont {Sch\"{o}nberger}, \citenamefont
  {de~Miranda~Cardoso}, \citenamefont {Reimer}, \citenamefont {Harrington},
  \citenamefont {Rodríguez}, \citenamefont {Nunez-Iglesias}, \citenamefont
  {Kuczynski}, \citenamefont {Tritz}, \citenamefont {Thoma}, \citenamefont
  {Newville}, \citenamefont {K\"{u}mmerer}, \citenamefont {Bolingbroke},
  \citenamefont {Tartre}, \citenamefont {Pak}, \citenamefont {Smith},
  \citenamefont {Nowaczyk}, \citenamefont {Shebanov}, \citenamefont {Pavlyk},
  \citenamefont {Brodtkorb}, \citenamefont {Lee}, \citenamefont {McGibbon},
  \citenamefont {Feldbauer}, \citenamefont {Lewis}, \citenamefont {Tygier},
  \citenamefont {Sievert}, \citenamefont {Vigna}, \citenamefont {Peterson},
  \citenamefont {More}, \citenamefont {Pudlik}, \citenamefont {Oshima},
  \citenamefont {Pingel}, \citenamefont {Robitaille}, \citenamefont {Spura},
  \citenamefont {Jones}, \citenamefont {Cera}, \citenamefont {Leslie},
  \citenamefont {Zito}, \citenamefont {Krauss}, \citenamefont {Upadhyay},
  \citenamefont {Halchenko},\ and\ \citenamefont
  {V\'{a}zquez-Baeza}}]{virtanen2020scipy}%
  \BibitemOpen
  \bibfield  {author} {\bibinfo {author} {\bibfnamefont {A.}~\bibnamefont
  {Vijaykumar}}, \bibinfo {author} {\bibfnamefont {A.~P.}\ \bibnamefont
  {Bardelli}}, \bibinfo {author} {\bibfnamefont {A.}~\bibnamefont {Rothberg}},
  \bibinfo {author} {\bibfnamefont {A.}~\bibnamefont {Hilboll}}, \bibinfo
  {author} {\bibfnamefont {A.}~\bibnamefont {Kloeckner}}, \bibinfo {author}
  {\bibfnamefont {A.}~\bibnamefont {Scopatz}}, \bibinfo {author} {\bibfnamefont
  {A.}~\bibnamefont {Lee}}, \bibinfo {author} {\bibfnamefont {A.}~\bibnamefont
  {Rokem}}, \bibinfo {author} {\bibfnamefont {C.~N.}\ \bibnamefont {Woods}},
  \bibinfo {author} {\bibfnamefont {C.}~\bibnamefont {Fulton}}, \bibinfo
  {author} {\bibfnamefont {C.}~\bibnamefont {Masson}}, \bibinfo {author}
  {\bibfnamefont {C.}~\bibnamefont {H\"{a}ggstr\"{o}m}}, \bibinfo {author}
  {\bibfnamefont {C.}~\bibnamefont {Fitzgerald}}, \bibinfo {author}
  {\bibfnamefont {D.~A.}\ \bibnamefont {Nicholson}}, \bibinfo {author}
  {\bibfnamefont {D.~R.}\ \bibnamefont {Hagen}}, \bibinfo {author}
  {\bibfnamefont {D.~V.}\ \bibnamefont {Pasechnik}}, \bibinfo {author}
  {\bibfnamefont {E.}~\bibnamefont {Olivetti}}, \bibinfo {author}
  {\bibfnamefont {E.}~\bibnamefont {Martin}}, \bibinfo {author} {\bibfnamefont
  {E.}~\bibnamefont {Wieser}}, \bibinfo {author} {\bibfnamefont
  {F.}~\bibnamefont {Silva}}, \bibinfo {author} {\bibfnamefont
  {F.}~\bibnamefont {Lenders}}, \bibinfo {author} {\bibfnamefont
  {F.}~\bibnamefont {Wilhelm}}, \bibinfo {author} {\bibfnamefont
  {G.}~\bibnamefont {Young}}, \bibinfo {author} {\bibfnamefont {G.~A.}\
  \bibnamefont {Price}}, \bibinfo {author} {\bibfnamefont {G.-L.}\ \bibnamefont
  {Ingold}}, \bibinfo {author} {\bibfnamefont {G.~E.}\ \bibnamefont {Allen}},
  \bibinfo {author} {\bibfnamefont {G.~R.}\ \bibnamefont {Lee}}, \bibinfo
  {author} {\bibfnamefont {H.}~\bibnamefont {Audren}}, \bibinfo {author}
  {\bibfnamefont {I.}~\bibnamefont {Probst}}, \bibinfo {author} {\bibfnamefont
  {J.~P.}\ \bibnamefont {Dietrich}}, \bibinfo {author} {\bibfnamefont
  {J.}~\bibnamefont {Silterra}}, \bibinfo {author} {\bibfnamefont {J.~T.}\
  \bibnamefont {Webber}}, \bibinfo {author} {\bibfnamefont {J.}~\bibnamefont
  {Slavi\v{c}}}, \bibinfo {author} {\bibfnamefont {J.}~\bibnamefont {Nothman}},
  \bibinfo {author} {\bibfnamefont {J.}~\bibnamefont {Buchner}}, \bibinfo
  {author} {\bibfnamefont {J.}~\bibnamefont {Kulick}}, \bibinfo {author}
  {\bibfnamefont {J.~L.}\ \bibnamefont {Sch\"{o}nberger}}, \bibinfo {author}
  {\bibfnamefont {J.~V.}\ \bibnamefont {de~Miranda~Cardoso}}, \bibinfo {author}
  {\bibfnamefont {J.}~\bibnamefont {Reimer}}, \bibinfo {author} {\bibfnamefont
  {J.}~\bibnamefont {Harrington}}, \bibinfo {author} {\bibfnamefont {J.~L.~C.}\
  \bibnamefont {Rodríguez}}, \bibinfo {author} {\bibfnamefont
  {J.}~\bibnamefont {Nunez-Iglesias}}, \bibinfo {author} {\bibfnamefont
  {J.}~\bibnamefont {Kuczynski}}, \bibinfo {author} {\bibfnamefont
  {K.}~\bibnamefont {Tritz}}, \bibinfo {author} {\bibfnamefont
  {M.}~\bibnamefont {Thoma}}, \bibinfo {author} {\bibfnamefont
  {M.}~\bibnamefont {Newville}}, \bibinfo {author} {\bibfnamefont
  {M.}~\bibnamefont {K\"{u}mmerer}}, \bibinfo {author} {\bibfnamefont
  {M.}~\bibnamefont {Bolingbroke}}, \bibinfo {author} {\bibfnamefont
  {M.}~\bibnamefont {Tartre}}, \bibinfo {author} {\bibfnamefont
  {M.}~\bibnamefont {Pak}}, \bibinfo {author} {\bibfnamefont {N.~J.}\
  \bibnamefont {Smith}}, \bibinfo {author} {\bibfnamefont {N.}~\bibnamefont
  {Nowaczyk}}, \bibinfo {author} {\bibfnamefont {N.}~\bibnamefont {Shebanov}},
  \bibinfo {author} {\bibfnamefont {O.}~\bibnamefont {Pavlyk}}, \bibinfo
  {author} {\bibfnamefont {P.~A.}\ \bibnamefont {Brodtkorb}}, \bibinfo {author}
  {\bibfnamefont {P.}~\bibnamefont {Lee}}, \bibinfo {author} {\bibfnamefont
  {R.~T.}\ \bibnamefont {McGibbon}}, \bibinfo {author} {\bibfnamefont
  {R.}~\bibnamefont {Feldbauer}}, \bibinfo {author} {\bibfnamefont
  {S.}~\bibnamefont {Lewis}}, \bibinfo {author} {\bibfnamefont
  {S.}~\bibnamefont {Tygier}}, \bibinfo {author} {\bibfnamefont
  {S.}~\bibnamefont {Sievert}}, \bibinfo {author} {\bibfnamefont
  {S.}~\bibnamefont {Vigna}}, \bibinfo {author} {\bibfnamefont
  {S.}~\bibnamefont {Peterson}}, \bibinfo {author} {\bibfnamefont
  {S.}~\bibnamefont {More}}, \bibinfo {author} {\bibfnamefont {T.}~\bibnamefont
  {Pudlik}}, \bibinfo {author} {\bibfnamefont {T.}~\bibnamefont {Oshima}},
  \bibinfo {author} {\bibfnamefont {T.~J.}\ \bibnamefont {Pingel}}, \bibinfo
  {author} {\bibfnamefont {T.~P.}\ \bibnamefont {Robitaille}}, \bibinfo
  {author} {\bibfnamefont {T.}~\bibnamefont {Spura}}, \bibinfo {author}
  {\bibfnamefont {T.~R.}\ \bibnamefont {Jones}}, \bibinfo {author}
  {\bibfnamefont {T.}~\bibnamefont {Cera}}, \bibinfo {author} {\bibfnamefont
  {T.}~\bibnamefont {Leslie}}, \bibinfo {author} {\bibfnamefont
  {T.}~\bibnamefont {Zito}}, \bibinfo {author} {\bibfnamefont {T.}~\bibnamefont
  {Krauss}}, \bibinfo {author} {\bibfnamefont {U.}~\bibnamefont {Upadhyay}},
  \bibinfo {author} {\bibfnamefont {Y.~O.}\ \bibnamefont {Halchenko}}, \ and\
  \bibinfo {author} {\bibfnamefont {Y.}~\bibnamefont {V\'{a}zquez-Baeza}},\
  }\href@noop {} {\bibfield  {journal} {\bibinfo  {journal} {Nat. Methods}\
  }\textbf {\bibinfo {volume} {17}},\ \bibinfo {pages} {261} (\bibinfo {year}
  {2020})}\BibitemShut {NoStop}%
\end{thebibliography}%

\clearpage
\onecolumngrid
\renewcommand\thefigure{S\arabic{figure}}
\renewcommand\theequation{S\arabic{equation}}
\renewcommand\thesection{S\arabic{section}}
\setcounter{figure}{0}
\setcounter{equation}{0}
\setcounter{section}{0}

\noindent {\Large \tbf{Supporting Information}}

\suppressfloats

\vspace{0.25cm}

\section{The effect of including $\langle V\rangle_0$ on $F_{\rm chg}^{\rm (Born)}$}

\begin{figure}[tb]
  \includegraphics[width=8cm]{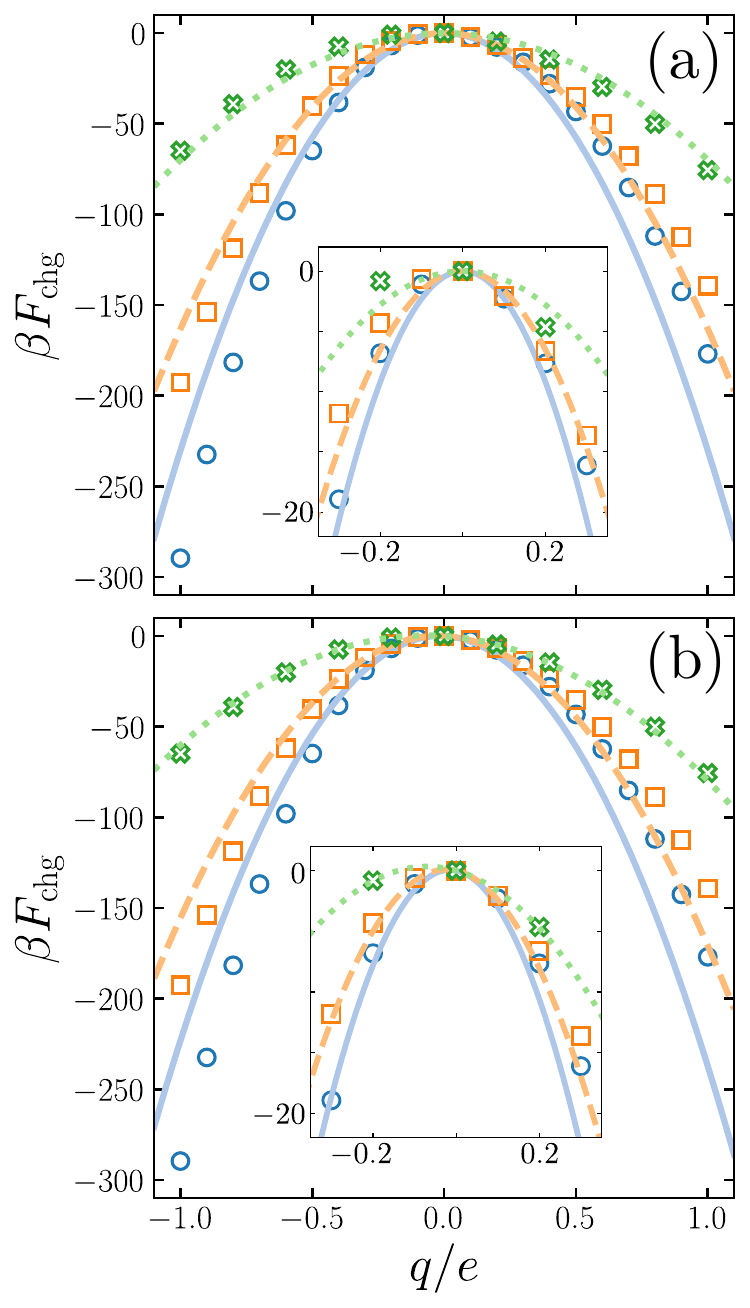}
  \caption{$F_{\rm chg}$ vs $q$ for different solute sizes $R_0$:
    circles, 2.70\,\AA; squares, 3.17\,\AA; crosses,
    5.50\,\AA. Symbols show results from simulations. (a) As shown in
    the main article, $F_{\rm chg}^{\rm (Born)}(q;0)$ \tcr{(Eq.~5)}
    largely captures the overall scale and size dependence of $F_{\rm
      chg}$, but it does not describe the asymmetric solvation of
    anions vs cations. Lines indicate best-fits of $F_{\rm chg}^{\rm
      (Born)}(q;0)$ to $F_{\rm chg}$. (b) Including a contribution
    $q\langle V\rangle_0$, $F_{\rm chg}^{\rm (Born)}(q;\langle
    V\rangle_0)$, only has a small effect, as seen by the similarity
    to panel (a). In both (a) and (b), the effective Born radii are
    found to be 1.21\,\AA{} (blue circles), 1.69\,\AA{} (orange
    squares) and 3.95\,\AA{} (green crosses). Insets: detailed view of
    the behavior for $-0.3\le q/e \le 0.3$ where the effect of adding
    $q\langle V\rangle_0$ is most clear.}
  \label{fig_si:fe_born}
\end{figure}

In Fig.~\ref{fig_si:fe_born} we present the results of including a term
$q\langle V\rangle_0$ in the Born model of solvation \tcr{(Eq.~5)}. On
the scale of Fig.~\ref{fig_si:fe_born}, this has negligible impact on
$F_{\rm chg}^{\rm (Born)}$. As seen in the insets of
Fig.~\ref{fig_si:fe_born}, the effect of including $q\langle V\rangle_0$
is most pronounced for small $q$.

\section{Contributions from more distant solvation shells}
\label{sec:shells}

\begin{figure}[tb]
  \includegraphics[width=7.65cm]{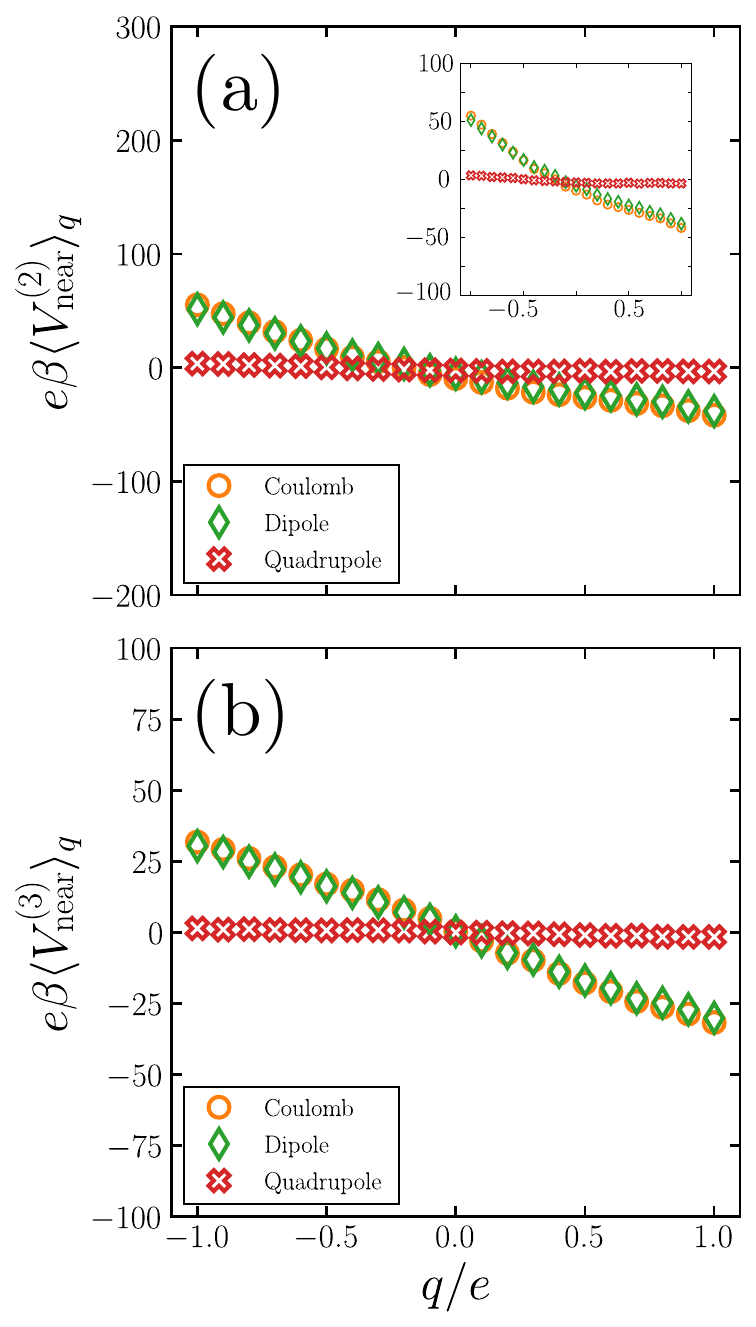}
  \caption{Contribution to $\langle V\rangle_q$ from more distant
    solvation shells. The degree on non-linearity is far less
    pronounced compared to the contributions from the first solvation
    shell (\tcr{Fig.~2}). (a) $\langle V_{\rm near}^{(2)}\rangle_q$ is
    the contribution from molecules in the second solvation
    shell. Inset: same data on a reduced scale, indicating there is
    still a degree of non-linear response. (b) $\langle V_{\rm
      near}^{(3)}\rangle_q$ is the contribution from molecules in the
    third solvation shell.}
  \label{fig_si:shells}
\end{figure}

Fig.~\ref{fig_si:shells} shows $\langle V_{\rm near}^{(2)}\rangle_q$ and
$\langle V_{\rm near}^{(3)}\rangle_q$, the contributions to $\langle
V\rangle_q$ from the second and third solvation shells, respectively,
for the same system shown in \tcr{Fig.~2} in the main text. Denoting
the distance between the center of the solute and the oxygen atom of a
water molecule as $R_{\rm SO}$, a molecule is deemed to be in the
second coordination shell if $3.5\,\text{\AA{}} < R_{\rm SO} \le
5.5$\,\AA, and within the third coordination shell if
$5.5\,\text{\AA{}} < R_{\rm SO} \le 8.5$\,\AA. The results indicate
that linear response is a reasonable approximation for the solvent's
dielectric response beyond the first solvation shell.

\section{Results for all solute sizes investigated}
\label{sec:allsizes}

\begin{figure}[tb]
  \includegraphics[width=16cm]{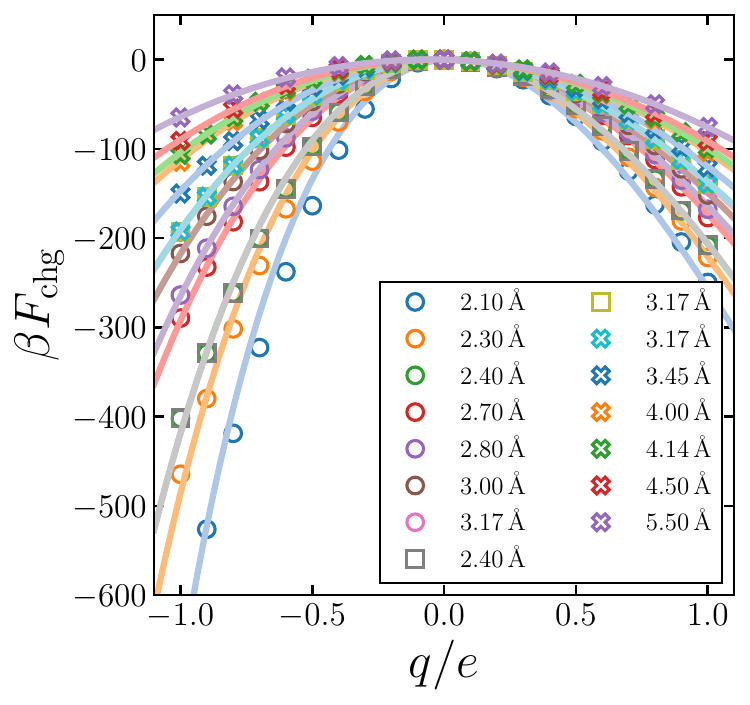}
  \caption{$F_{\rm chg}$ for all solute sizes studied. Circles,
    squares and crosses indicate data obtained from simulations with
    64, 256 and 512 water molecules, respectively. Results for $R_0 =
    2.40$\,\AA{} have been obtained with both 64 and 256 water
    molecules, while for $R_0 = 3.17$\,\AA, $F_{\rm chg}$ has been
    computed for all three system sizes: these results indicate that
    the finite size corrections work as expected. Solid lines indicate
    best fits of $F_{\rm chg}^{\rm (var)}$ to the simulation
    data. While some discrepancies are observed for the smallest
    solutes, the theory does a reasonable job at capturing the charge
    asymmetry.}
  \label{fig_si:allsizes}
\end{figure}

In Fig.~\ref{fig_si:allsizes} we show $F^{\rm (var)}_{\rm chg}$ fitted to
$F_{\rm chg}$ obtained from simulation for all solute sizes
investigated. For the smallest solutes we see some relatively small
discrepancies between the simulation and the theory, but the large
degree of charge asymmetry is nevertheless captured. Also shown in
Fig.~\ref{fig_si:allsizes} are results for the same solute in different
sized simulation boxes, indicating the finite size corrections
described in the main text are sufficient to obtain an estimate for
the macroscopic charging free energies.

\section{Dielectric radii}
\label{sec:Reff}

\begin{figure}[tb]
  \includegraphics[width=8cm]{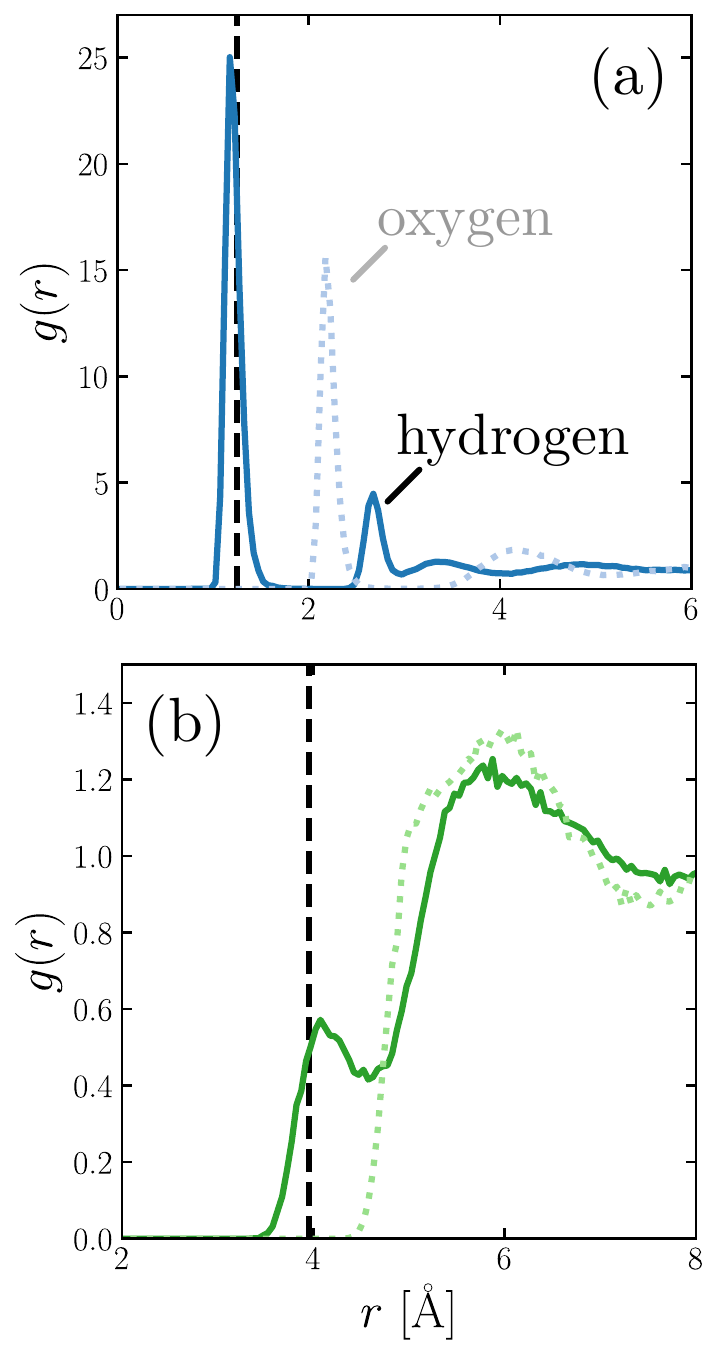}
  \caption{Solute-solvent radial distribution functions for (a) $R_0 =
    2.70$\,\AA{} and (b) $R_0 = 5.50$\,\AA. In both cases
    $q=-e$. Solid and dotted lines show solute-hydrogen $g(r)$ and
    solute-oxygen $g(r)$, respectively. The vertical dashed line
    indicates $R$.}
  \label{fig_si:rdfs}
\end{figure}

In Fig.~\ref{fig_si:rdfs} we show a plot analogous to \tcr{Fig.~4b} in
the main article, but for solute sizes $R_0=2.70$\,\AA{} and
$R_0=5.50$\,\AA. In both cases we find that $R$ roughly corresponds to
the distance of closest approach for the hydrogen atoms of the water
molecules.

\section{Equipotential surfaces}

\begin{figure}[tb]
  \includegraphics[width=8cm]{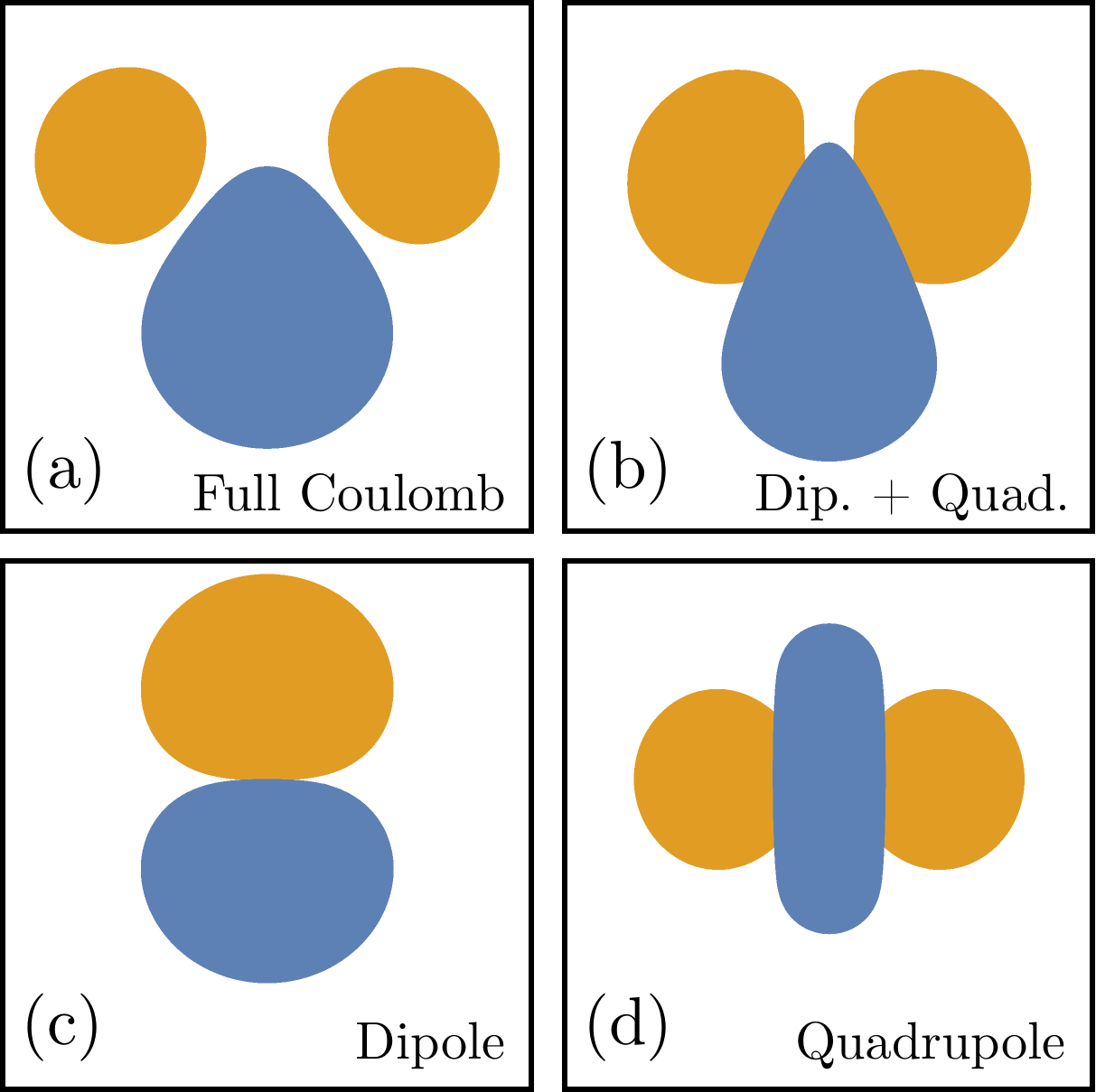}
  \caption{(a) Equipotential surface of a single water molecule. (b)
    Equipotential surface resulting from (c) dipole and (d) quadrupole
    contributions. The center of charge has been used for the
    multipole expansion.}
  \label{fig_si:EquiPot}
\end{figure}

\begin{figure}[tb]
  \includegraphics[width=8cm]{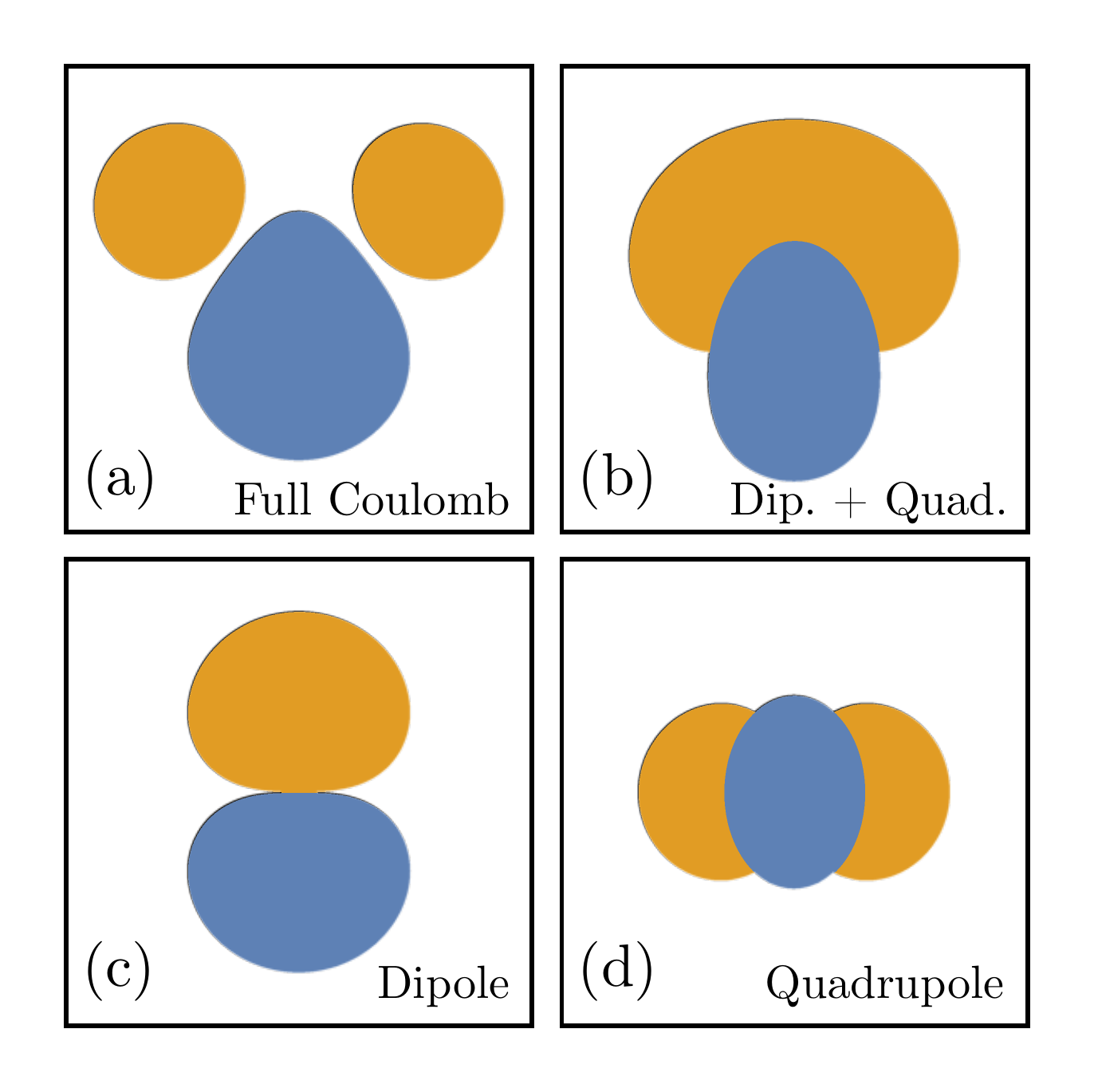}
  \caption{(a) Equipotential surface of a single water molecule. (b)
    Equipotential surface resulting from (c) dipole and (d) quadrupole
    contributions. The position of the oxygen atom has been used for
    the multipole expansion.}
  \label{fig_si:EquiPotOxy}
\end{figure}

Figures~\ref{fig_si:EquiPot} and~\ref{fig_si:EquiPotOxy} show equipotential
surfaces arising from dipole and quadrupole contributions, with the
multipole expansion respectively performed around the center of
charge, and the position of the oxygen atom. Using the center of
charge results in an equipotential surface that more closely resembles
that of SPC/E water.

\end{document}